\documentclass[aps,pre,twocolumn,showpacs,groupedaddress]{revtex4-2}

\usepackage{graphicx,textcomp,color,bm,mismath,comment}

\def \veps {{\large{\boldsymbol{\varepsilon}}}}

\begin{document}

\title{A device for studying elementary plasticity fluctuations in granular media}

\author{Ambroise Mathey}
\author{Micka\"el Le Fur}
\author{Patrick Chasle}
\author{Axelle Amon}
\author{J\'er\^ome Crassous} \affiliation{Univ Rennes, CNRS, IPR (Institut de Physique de Rennes) - UMR 6251, F-35000 Rennes, France}
\email{jerome.crassous@univ-rennes1.fr}

\date{\today}

\begin{abstract}
{In this manuscript, we describe a scientific device specifically designed for the study of the plasticity fluctuations preceding the fracture of granular media. Biaxial tests on model granular media are performed using a commercial uniaxial loading system. Strain field fluctuations are measured using a method based on the interference of coherent light scattered by the sample. We show that such a device enables discrete plasticity events to be unambiguously evidenced. Moreover, those discrete plasticity fluctuations depend only on the imposed strain, and not on the strain rate.}
\end{abstract}

\maketitle

%%%%%%%%%%%%%%%%%%%%%%%%%%%%%%%%%%%%%%%%%%%%%%%%%%%%%%%%%%%%%%%%%%%%%%%
\section{Introduction}

The plasticity and fracture of granular media is an interesting subject for many reasons. From a practical point of view, mechanics seek to understand the resistance that a medium (sand, aggregate, natural soil, etc.) can undergo without or with negligible deformation~\cite{nedderman,davis}. This is of crucial importance for structural design. From a more fundamental point of view, describing the flow of these materials from a static state is a theoretical challenge that is still being debated~\cite{andreotti.book,alexandre.2018}. Experimentally and macroscopically, the behavior of these materials is relatively clear and universal. For dry granular materials, subjected to homogeneous shear stresses, the materials exhibit elastic behavior for very low strains, followed by relatively homogeneous plastic behavior~\cite{bardet1990,desrues2004,davis,viggiani2012}. At higher deviatoric stresses, a deformation localization mode of failure occurs, usually with the appearance of shear bands in the system~\cite{desrues2004,vardoulakis}. More complex deformation modes -diffuse localization~\cite{darve.2013}, compaction bands~\cite{besuelle2004}, intermittent deformation~\cite{lherminier2019}, etc.- can also occur, depending on the initial state of preparation of the system.

From a fundamental point of view, these quasi-statically deformed granular systems are both athermal and disordered, and thus belong to the class of glassy systems. The description of glassy systems by statistical physics has made great advances recently, particularly in the flow of materials such as concentrated colloidal systems, pastes, gels and so on~\cite{alexandre.2018}. In particular, the rheology of these materials relies on their ability to reorganize on a microscopic scale in the form of local plasticity events~\cite{argon1979,falk.1998, picard2004,alexandre.2018}, These events can be coupled via the elasticity of the materials, giving rise to localised flow structures~\cite{falk.1998,bocquet.2009,Goyon2008}. It should be noted that these results were obtained for system with no microscopic friction. In the case of granular media, it is not then obvious - {\it a priori} - that such localised deformations may occur in this theoretical framework. A different point of view is to consider the deformation in granular material as the result of deformation modes in sheared material. The approach of the failure of the materials occurs with deformation modes that progressively span across the system~\cite{lin.2015}. In the limit of non-compressible materials, it has been argued, based on numerical simulations, that localised deformation can not occur in sheared granular systems~\cite{Bouzid.2015}. This last statement does not agree with results of numerical simulations that show that localised deformation occur in the flow of rigid spheres~\cite{Liu.2022,liu.2022a,zheng2023micromechanical,ma.2021}. 

Finally, the interactions between such local reorganisations inside a frictional granular material is not clear. First there is some experiments~\cite{lebouil2014,houdoux2018} and numerical simulations~\cite{kuhn1999, guo2014, mcnamara.2016, darve2021, Wang2022} that transient positive correlations of stress or strain occurs along fixed (respectively to the applied stress axes) directions. Those directions are in agreement with the existence of elastic coupling that are given by Eshelby tensor~\cite{eshelby1957}, even in the presence of microscopic friction~\cite{Karimi2018}. Their spatial extent can be assumed to be a few grains by analogy with numerical studies of the rheology of glassy systems, but without clear experimental evidence for granular media. The amplitude of these events (e.g. strain or stress release), and the statistical properties of these events are unknown.  Finally, it is unclear whether the nature of such events is the same homogeneous plastic deformation (i.e. before deformation localization) and in a phase of localized flow are the same.

The aim of this study is to present a device specifically designed for the experimental observation of localized deformations occurring in granular materials under deviatoric stress. The mechanical principle of the test is standard (biaxial compression) and is coupled to an inferential displacement method developed previously. The specific features of this new apparatus compared with a previous study~\cite{lebouil.2014a} are twofold. Firstly, the mechanical test is carried out on a commercial uniaxial tension/compression machine. This has the advantage of being easily reproducible by the interested community. Also, wide ranges of compression speeds can be easily explored, with in particular very constant strain rates, even at very low strain rates. Secondly, the acquisition of interference patterns (from which local strain measurements are derived) is carried out in such a way as to be able to measure small fluctuations in strain, even in the presence of optical signal drift. The combination of these two points makes it possible to test the dependence of statistical properties of these events on strain rates.

The manuscript is organised as follow. In section~\ref{sec:2} we present the mechanical part of the setup. In section~\ref{sec:3} we discuss the principle of the local deformation measurement using light scattering technique. We then discuss a typical experiment in section~\ref{sec:4}. Preliminary results on the statistic of those events are finally presented in  section~\ref{sec:5}.

%%%%%%%%%%%%%%%%%%%%%%%%%%%%%%%%%%%%%%%%%%%%%%%%%%%%%%%%%%%%%%%%%%%%%%%
\section{Mechanical system}\label{sec:2}
\subsection{Overview}

\begin{figure}
\centering
\includegraphics[width=\columnwidth]{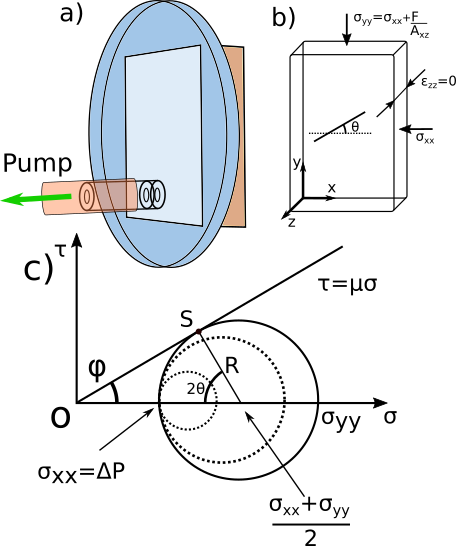}
\caption{a) Sample representation b) Biaxial test principle. c) Mohr-Coulomb failure analysis: The experiment starts with $\sigma_{xx}=\sigma_{yy}=\Delta P$. $\sigma_{yy}$ is then increased until Mohr's circle touches the Coulomb line $\tau=\mu \sigma$, with $\tau$ (resp. $\sigma$) the tangential (resp. normal) stress along a failure plane, and $\mu$ the internal friction coefficient. The point $S$ correspond to formation of shear band in the system.}\label{fig:mohr_coulomb}
\end{figure}

The apparatus is designed to perform a  biaxial and homogeneous stress test on a sample, with the ability to see one face of the sample through one glass plate for acquisition of images of interference. For this, a granular sample is placed inside rectangular cuboid shape made of an elastic membrane on 5 sides, and one glass plate on the last side. The sample is confined between a fixed glass plate and 4 metallic plate : one is parallel to the glass plate, and two perpendicular to the compression axis. We ensure that there is no deformation of those two plates, and then that the displacement and deformation  in the $z$-direction is null: $\varepsilon_{zz}=0$. A difference of pressure $\Delta P=P_{out}-P_{in}$, with $P_{out}$ the external atmospheric pressure, and $P_{in}$ the pressure within the elastic membrane is applied, creating an homogeneous confining lateral stress $\sigma_{xx}$. Finally, the sample is compressed along $y$ direction, and the stress $\sigma_{yy}$ is obtained as $\sigma_{yy}=\sigma_{xx}+F/A_{xz}$, where $F$ is the applied compression force, and $A_{xz}$ the area of the sample in a plane normal to $y$. Experiments are performed at fixed $\Delta P=\sigma_{xx}$. Initially, $F\simeq 0$, and $\sigma_{yy} \simeq \sigma_{xx}$, and the top plate is moved at constant velocity. During the compression, $\sigma_{yy}$ increases until the rupture occurs. Then, the material flows at a roughly constant value of $\sigma_{yy}$ with one or two shear bands.
     
\subsection{Translation Device} 

 To compress the sample, we use a commercial uni-axial loading test machine. Having a commercial device allows for easy utilisation and well-tested reliability while allowing other research teams to reproduce the experiment. The machine is an Instron 5965 Dual Column Table Frame with a $5~kN$ load capacity. This limit is enforced by a load cell (2580 Series $5~kN$ Static Load Cell from Instron) that stops the system from applying more force. This force limit also gives us a maximal size for the sample under study. Indeed, through Mohr-Coulomb failure analysis, a failure plane can form if the shear stress $\tau$ and the normal stress $\sigma$ acting on the plane follow (for an ideal cohesion-less Coulomb material) the Coulomb yield criterion: $\tau=\mu \sigma$ as represented by the straight line in fig.~\ref{fig:mohr_coulomb}, with $\mu$ the internal friction coefficient. The normal and shear stresses at failure may be expressed as a function of the applied principal stresses~\cite{nedderman}: $\sigma=p+R\cos{2\theta}$ and $\tau=R\sin{2\theta}$ with $p=(\sigma_{xx}+\sigma_{yy})/2$, $R=(\sigma_{yy}-\sigma_{yy})/2$, and $\theta$ the angle between shear plane and $x$ axis (see fig.~\ref{fig:mohr_coulomb}.b), and finally $(\sigma_{yy}-\sigma_{xx})/(\sigma_{yy}+\sigma_{xx})=sin(\varphi)$, where $\mu=\tan(\varphi)$. With $\sigma_{yy}=\sigma_{xx}+F/A_{xz}$, and $\sigma_{xx}=\Delta P$, we obtain $F= A_{xz}~\Delta P~2\sin\varphi/(1-\sin\varphi)$ at failure. For maximal values $\mu \simeq 1$, $F=5~kN$,  $\Delta P=10^5~Pa$, we obtain that the section $A_{xz}$ must be smaller than $10^{-2}~m^2$. So in practice any sample of decametric size may be studied with the maximal pressure on confinement that we use. The compression force and the displacement are obtained through the proprietary software Bluehill Universal 4.16.
 
 An important point for our applications is that the displacement of the loading part is smooth and does not show any discrete steps. We checked this point by measuring the displacement of the translating part of the Instron  using an inductive sensor. The result is that in a translation velocity range $\bigl[0.1 \mu m .s^{-1};5 ~cm.s^{-1}\bigr]$, no steps are visible (i.e. steps, if present are smaller than $44 nm$).

%%%%%%%%%%%%%%%%%%%%%%%%%%%%%%%%%%%%%%%%%%%%%%%%%%%%%%
\subsection{Sample mounting system}   
\begin{figure*}[ht!]
\centering
\includegraphics[width=0.7\paperwidth]{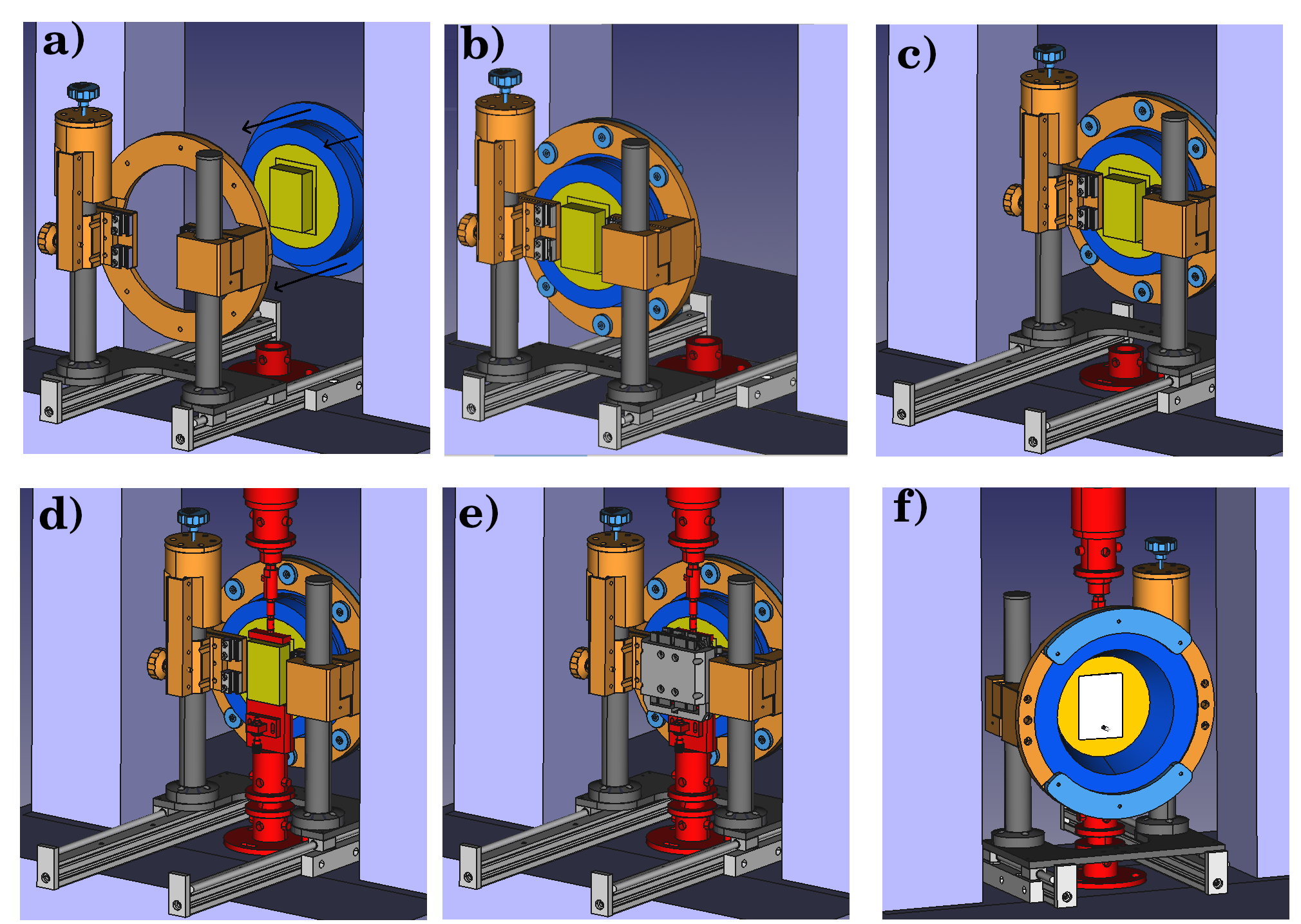}
\caption{Instron machine(light purple) and mounting system(orange): The sample (yellow), the glass panel's rim (blue), mounting screws (light blue), the rail (light grey), the Instron's support and compression parts (red) and the adjustable back plate (black and grey)}
\label{fig:instron_mounting_sys}
\end{figure*}
 
As shown in fig.~\ref{fig:instron_mounting_sys} the sample and the glass panel's rim are placed in the mounting system (a) before being secured into place using mounting screws (b), allowing the whole system to slide into place using the rails (c). The Instron's support and compression parts are then put in place (d) along with the adjustable back plate which is pressed against the back of the sample (e). It is now ready to be submitted to a biaxial test and imaged through the glass panel (f).

%%%%%%%%%%%%%%%%%%%%%%%%%%%%%%%%%%%%%%%%%%%%%%%%%%%%%%%%%    
\section{Local deformation measurement}\label{sec:3}
The non-conventional imaging technique used here is based on interference in scattering medium. In section~\ref{sec:3.1} we just present main theoretical results, the details may be found in previously published papers~\cite{djaoui.2005,crassous.2007, erpelding.2010, amon.2017}. The section~\ref{sec:3.2} is devoted to the experimental determination of intensity correlation functions and is specific (camera, framerate, level of noise) to the set-up and to the measurements made here.

%%%%%%%%%%%%%%%%%%%%%%%%%%%%%%%%%%%%%%%%%%%%%%%%%%%%%%%%%
\subsection{Interferometric method.}\label{sec:3.1}
\subsubsection{Intensity correlation functions.}

\begin{figure}[ht!]
\centering
\includegraphics[width=\columnwidth]{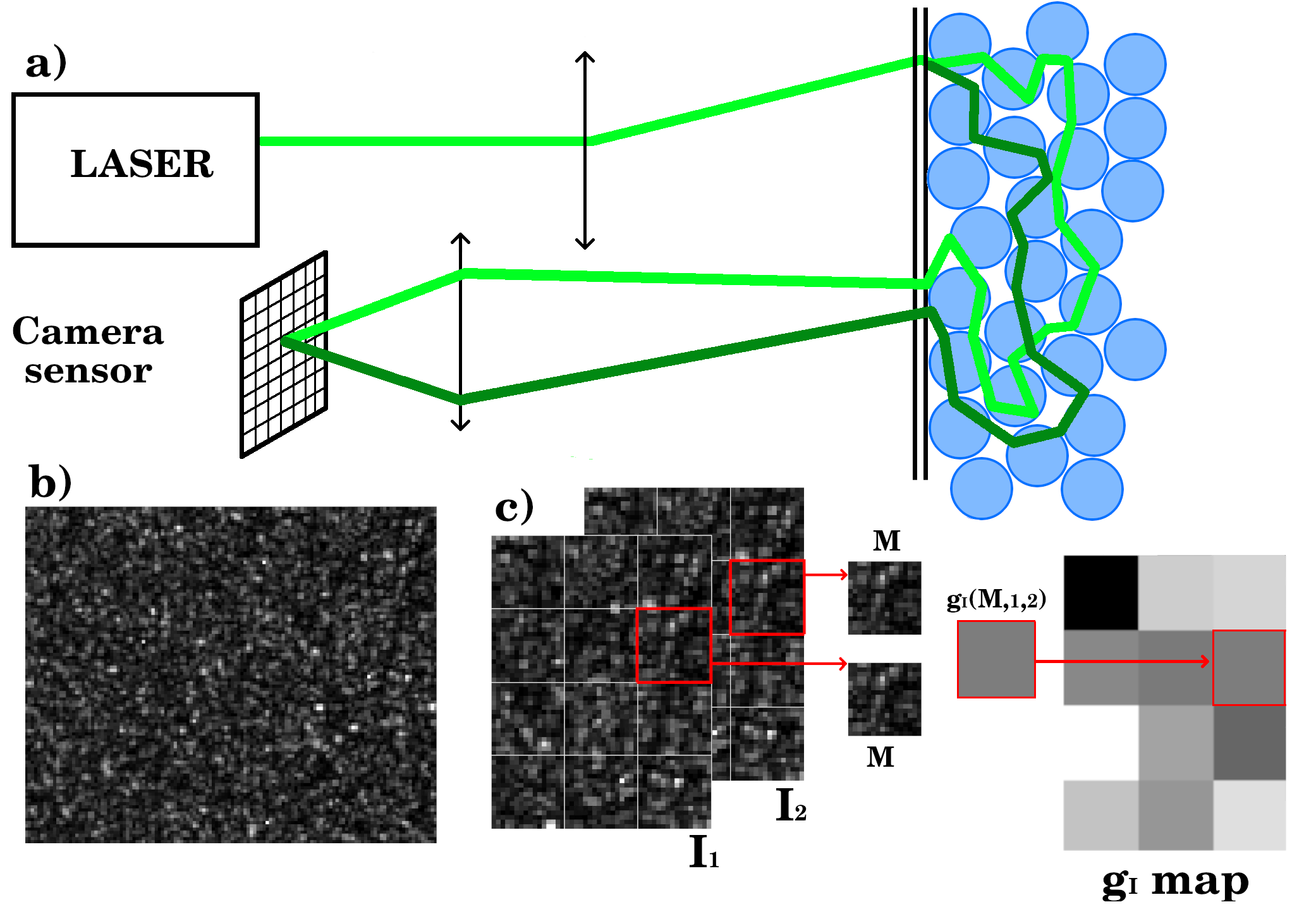}
\caption{a) Principle of light scattering measurement. b) Speckle pattern recorded with a camera c) Computation of the $g_I(M,1,2)$ map}
\label{fig:DWS}
\end{figure}

Interference in a highly scattering medium may be described using Diffusive Wave Scattering (DWS) formalism~\cite{pine.1990,pine.1990b}. A laser beam illuminates a scattering medium (here our granular matxxerial) and, because of scattering, different optical paths are followed in the material (see fig.\ref{fig:DWS}.a). Those different paths interfere on a camera sensor. We use the backscattering geometry where the scattered intensity is measured on the side that is illuminated with the camera focused on the same face. We obtain speckle images (see fig.\ref{fig:DWS}.b) that are the results of the interference between the different optical paths. The size of these speckle spots is dependent on the geometry of illumination and detection, and not on the nature of the granular material under study. The size of the speckle spots on the camera needs to meet 2 requirements: having enough speckle patterns on our images to have good statistics for the determination of the intensity correlation functions, and the speckle spots should be larger than one pixel of the camera. Following~\cite{viasnoff.2002}, a speckle size of $2$ pixels is the best compromise.

To identify the displacement of the grains, we consider two different speckle images (here called $1$ and $2$) taken for 2 arbitrary deformation states. Each pixel $p$ has received a certain light intensity $I_1(p)$ or $I_2(p)$. We divide the image into square groups of $16\times 16$ pixels, which we call metapixel, and we calculate the correlation function of the scattered intensity for every metapixel $M$:

\begin{equation}
g_I(M,1,2)=\frac{\langle I_1 I_2 \rangle-\langle I_1 \rangle \langle I_2 \rangle}{\sqrt{\langle I_1^2 \rangle-\langle I_1 \rangle^2}
\sqrt{\langle I_2^2 \rangle-\langle I_2 \rangle^2}}\label{eq:defgI}
\end{equation}

where $\langle f \rangle = \frac{1}{N_p}\sum \limits_{p \in M}f(p)$, with $N_p$ the number of pixel in the metapixel $M$, is the mean value of an arbitrary quantity $f(p)$ on the metapixel $M$. The practical way to compute this quantity in the presence of electronic noise is devoted to section~\ref{sec:noise_reduction}. By construction, for any state $i=1,2,..$, we have  $g_I(M,i,i)=1$ for every metapixel $M$.
            
For a couple of states $1$ and $2$, the picture made of metapixels with gray values proportional to $g_I(M,1,2)$ represents a map showing where the scattered light is correlated (see fig.~\ref{fig:DWS}.c).

%%%%%%%%%%%%%%%%%%%%%%%%%%%%%%%%%%%%%%%%%%%%%%%%%%%%%%% 
\subsubsection{Relating $g_I$ to the local deformation.}

Obtaining quantitative information about the beads' displacement in a metapixel from its $g_I$ value requires an optic model~\cite{erpelding.2010,amon.2017}. This model does not give access to the full deformation tensor $\veps$ in the material, but only to an isotropic invariant of the deformation tensor. The hypotheses of the model are the following: i) The displacement field of the grains between state $1$ and $2$ is locally affine. ii) The deformation tensor is small and homogeneous in the material at the scale of the metapixel. iii) The scattering light is collected in backscattering geometry. Under those hypotheses~\cite{erpelding.2010,amon.2017}:
\begin{equation}
g_I(M,1,2)\approx \exp\left(-2\eta kl^* \sqrt{3f(\veps)}\right)
\end{equation}
with
\begin{equation}
f(\veps)=\frac{1}{15}Tr^2(\veps)+\frac{2}{15}Tr(\veps^2)
\end{equation}
where $l^*$ is the transport mean free path (this length, related to the optical properties of the material, is typically few grain size for granular material), $k=2\pi/\lambda$ (with $\lambda$ the laser wavelength) is the wave number, and $\eta$ an optical factor depending of boundaries conditions for light, that is typically $\eta \approx 1-2$.

Different kinds of deformation may be probed. If we assume that the local deformation tensor is coaxial to the imposed stress, with no expansion along $z$-axis, and isovolume, we expect a local deformation tensor:
\begin{equation}
\veps=\left[\begin{array}{cccc}
\varepsilon_{xx}  & 0 & 0 \\
0 & -\varepsilon_{xx} & 0  \\
0 & 0 & 0 
\end{array}\right]
\end{equation}
In this case, $f(\veps)=4\varepsilon_{xx}/15$, and:
\begin{equation}
 g_I\approx \exp \left(-2\eta kl^* \sqrt{\frac{4}{5}\varepsilon_{xx}^2} \right)
\end{equation}
Or equivalently
\begin{equation}
\varepsilon_{xx}\approx\frac{-ln(g_I)}{C}
\end{equation}
with $C=\frac{4}{\sqrt{5}}\eta k l^*$. This value is $C \approx 3.8 \times 10^{3}$, for grains of 90 $\mu m$ in diameter.
One of the interests of this method of measurement despite its complexity is its sensitivity. The typical local deformation $\varepsilon_{xx}$ we measure in an event is between $10^{-4}$ and $10^{-6}$. It also provides some spatial resolution for these motions. Indeed we can tell inside which metapixel it happened (1 metapixel corresponds to a square of 500$\mu m$ by 500$\mu m$ and contains roughly 25 grains). This is a great improvement over the acoustic methods where very small deformations may be easily recorded but with no information on their localisation. A drawback of those methods is that the zone of the material that is probed is only the illuminated one. In backscattering geometry, the light penetrates the sample on a typical length (see~\cite{pine.1990b} for details) of $5-10~d$, and then we only probe the material at its surface.

%%%%%%%%%%%%%%%%%%%%%%%%%%%%%%%%%%%%%%%%%%%%%%%%%%%%%%%%%%%%%%%%
\subsection{Experimental measurement of $g_I$}\label{sec:3.2}

This section will discuss the specifics of the measurement of the correlation function $g_I$ in our experimental setup.

\subsubsection{Laser and Camera}
To illuminate the whole sample uniformly, we use a Coherent Laser (Compass 215M-75 Coherent) with a wavelength $\lambda=532$ nm and an output power of 75 mW that is broadened using a lens of focal length 8mm to illuminated all the sample. The imaging is then done using a DALSA camera (Falcon-4M60, resolution 2352 x 1728, and size 7.4 $\mu m$, 8 bits) after passing through an aperture, a monochromatic filter at $\lambda$ and a lens of focal length 100mm. The large resolution of the camera allows our metapixel to be large and therefore contain the information of a large amount of pixels thus reducing fluctuations in the pixels. However, as the metapixels get bigger it reduces the spatial resolution mentioned before. We settled for metapixels of 16 by 16 pixels as a good in-between~\cite{erpelding.2010,viasnoff.2002,aime.2018}.

\subsubsection{Camera noise correction}\label{sec:noise_reduction}
In our experiments, we are interested in increasing our frame rate and therefore reducing the amount of light our camera's sensor receives. This increases the significance of the electric noise of the pixels on our signal. The correlation of the electronic noise must be estimated and removed from the raw correlation function to obtain $g_I$. We consider here images recorded at a given duration of exposure $\tau_{exp}$. Let's $I_n(p,t)$ the electronic noise for the image recorded at time $t$, and $I_s(p,t)$ the signal of intensity that we record. The total (measured) intensity is:
\begin{equation}
I_m(p,t)=I_s(p,t)+I_n(p,t)
\label{eq:Im}
\end{equation}

We do the hypothesis that the electronic noise is uncorrelated to the signal noise at the scale of one metapixel $M$:
\begin{equation}
\langle I_s(p,t) I_n(p,t') \rangle=\langle I_s(p,t) \rangle~\langle I_n(p,t') \rangle
\label{eq:IsIp}
\end{equation}
$\forall t,\forall t',\forall M$.

We write the noise at each pixel as:
\begin{equation}
I_n(p,t)=I_n^{(av)}(p)+I_n^{(sd)}(p)\xi_p(t)
\label{eq:Inga}
\end{equation}
where $I_n^{(av)}(p)$ is the average value and $I_n^{(sd)}(p)$ the standard deviation of the noise of pixel $p$, (i.e. $\overline{\xi_p(t)}=0$ and $\overline{\xi^2_p(t)}=1$ with $\overline{f}$ the time average of a quantity $f$). We \textcolor{red}{assume} that the noise of the camera verifies:
\begin{equation}
\overline{ \xi_p(t)~\xi_{p'}(t+\tau)}=\delta(p'-p)~\delta(\tau)
\label{eq:xixi}
\end{equation}
the noise at different pixels are uncorrelated, and the noise at one pixel is $\delta$-correlated in delay $\tau$.

We note in the following:
\begin{align}
G_s(M,t,t+\tau)
&=\langle I_s(p,t) I_s(p,t+\tau) \rangle\nonumber\\
&-\langle I_s(p,t) \rangle \langle I_s(p,t+\tau) \rangle\\
G_m(M,t,t+\tau)
&=\langle I_m(p,t) I_m(p,t+\tau) \rangle\nonumber\\
&-\langle I_m(p,t) \rangle \langle I_m(p,t+\tau) \rangle\\
G_n(M,t,t+\tau)
&=\langle I_n(p,t) I_n(p,t+\tau) \rangle\nonumber\\
&-\langle I_n(p,t) \rangle \langle I_n(p,t+\tau) \rangle
\end{align}
From the assumed independence between signal and noise, we have:
\begin{equation}
G_s(M,t,t+\tau)=G_m(M,t,t+\tau)-G_n(M,t,t+\tau)
\end{equation}
We split $G_n(M,t,t+\tau)$ in time-average and a varying parts:
\begin{align}
G_n(M,t,t+\tau)=\overline{G_n(M,t,t+\tau)}+\delta G_n(M,t,t+\tau)
\end{align}
where $\overline{G_n(M,t,t+\tau)}$ and $\delta G_n(M,t,t+\tau)$ may be expressed using $I_n^{(av)}$, $I_n^{(sd)}$ and $\xi$ as:
\begin{align}
\overline{G_n(M,t,t+\tau)}&=
\langle I_n^{(av)}(p)^2 \rangle \delta(\tau)\nonumber\\
&+ \langle I_n^{(sd)}(p)^2 \rangle
- \langle I_n^{(sd)}(p)\rangle^2\label{eq:GnBar}
\end{align}
and
\begin{align}
&\delta G_n(M,t,t+\tau)=\nonumber\\
&\langle I_n^{(sd)}(p)^2 \rangle \bigl[ \langle \xi_p(t)\xi_p(t+\tau) \rangle-\delta(\tau)\bigr]\nonumber\\
+&\bigl[\langle I_n^{(av)}(p)I_n^{(sd)}(p) \rangle-
\langle I_n^{(av)}(p)\rangle \langle I_n^{(sd)}(p) \rangle\bigr]\nonumber\\
&\quad \bigl[\langle \xi_p(t) \rangle + \langle \xi_p(t+\tau) \rangle\bigr]\nonumber\\
-&\langle I_n^{(sd)}(p)^2 \rangle\  \langle \xi_p(t) \rangle  \langle \xi_p(t+\tau) \rangle
\end{align}
Finally
\begin{align}
G_s(M,t,t+\tau)&=G_m(M,t,t+\tau)\nonumber\\
&-\overline{G_n(M,t,t+\tau)}\nonumber\\
&-\delta G_n(M,t,t+\tau)
\end{align}
For each metapixel $M$, the value $\overline{G_n(M,t,t+\tau)}$ is obtained from a long measurement with no illumination. We found, in agreement with~\eqref{eq:xixi} that the noise is $\delta$-correlated in $\tau$, and we obtain the terms $\langle I_n^{(av)}(p)^2 \rangle$ and $\langle I_n^{(sd)}(p)^2 \rangle
- \langle I_n^{(sd)}(p)\rangle^2$ involved in~\eqref{eq:GnBar}. Those values are dependent of the exposure time, and do not drift on the scale of few weeks.

The fluctuating part $\delta G_n(M,t,t+\tau)$ is the contribution of the electronic noise to determination of the signal correlation function $G_s$. The typical magnitude of this contribution depends of the statistical properties of $I_n^{(av)}$, $I_n^{(sd)}$ and $\xi_p(t)$. By definition $\overline{\xi^2_p(t)}=1$, so that we may expect that $\vert \langle \xi_p(t) \rangle \vert\sim 1/\sqrt{N_p}$ and we can measure $\delta G_n(M,t,t+\tau)$ by calculating the standard variation of ${G_n(M,t,t+\tau)}$, giving $\vert\delta G_n(M,t,t+\tau)\vert\sim 0.056$. For typical values of signal intensity $I_s\sim80$ and correlation functions $G_s\sim 120$, we have $\vert\delta G_n(M,t,t+\tau)\vert \ll G_s$, so that we can write : 
\begin{align}
G_s(M,t,t+\tau) \simeq & ~G_m(M,t,t+\tau)\nonumber\\&-\overline{G_n(M,t,t+\tau)}
\end{align}
Defining $1$ (respectively $2$) the state of the system at time $t$ (resp. $t+\tau$), we obtained $g_I(M,1,2)$ as:
\begin{equation}
g_I(M,1,2)=\frac{G_s(M,t,t+\tau)}{\sqrt{G_s(M,t,t)}\sqrt{G_s(M,t+\tau,t+\tau)}}\label{eq:gI.exp}
\end{equation}

%%%%%%%%%%%%%%%%%%%%%%%%%%%%%%%%%%%%%%%%%%%%%%%%%%%%%%%
\subsection{Dynamical range of acquisition rate}

\subsubsection{High acquisition rate}
The upper limit of the acquisition rate of our optical detection is determined by hardware limitations. The first limit is the decrease in exposure time as the acquisition rate increases. The illumination diminishes, decreasing the signal-over-noise ratio. In practice, when the frame rate is $fps=50~Hz$, the correlation function of a static object (which theoretically should be $g_I(M,1,2)=1$, $\forall 1,2,M$) is lower than $1$ is typically $\vert g_I(M,1,2)-1\vert \sim 2\times10^{-3}$ for $t_2-t_1=1/fps$. This decay of correlation functions may be decreased by increasing the size $N_p$ of metapixel, but with a diminution of the spatial resolution of the image. More powerful lasers may also be used. The second limit is the writing capability of our SSD drives which were not able to write steadily at more than 80 MB/s (4 MB images at 20 fps) without reaching the limited writing speeds of disks. We were able to go faster (40 fps) by reducing the resolution of our speckle images to 1 MB.

\subsubsection{Low acquisition rate}\label{sec:low_acqui_rate}
The low limit of the acquisition rate is fixed by the long-time stability of our optical system. As the deformation increments and therefore the time between images increases, some lack of stability in our setup are brought out as we observe decorrelation on motionless objects. We suspect that these decorrelations are caused by optical mode fluctuations in our laser caused by its heating over extended use. It could also be caused by the convection of air in our experiment. While the attempt at identifying those decorrelations has been unsuccessful, a good signal-to-noise ratio reduces their significance.

%%%%%%%%%%%%%%%%%%%%%%%%%%%%%%%%%%%%%%%%%%%%%%%%%%%%%%%
\section{Observation of the fluctuation of the local deformation}~\label{sec:4}

%%%%%%%%%%%%%%%%%%%%%%%%%%%%%%%%%%%%%%%%%%%%%%%%%%%%%%%
\subsection{Sample preparation}
\begin{figure}[ht!]\centering\includegraphics[width=\columnwidth]{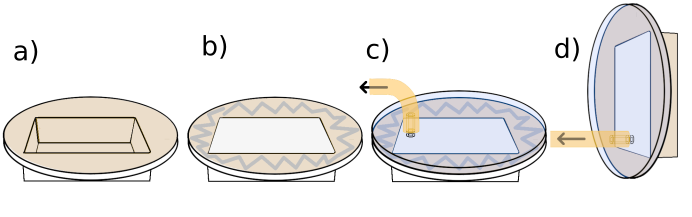}
\caption{ Sample preparation: a) Placing the latex membrane in a mould. b) Filling the membrane with dry beads and applying vacuum grease on the side of the membrane. c) Placing the glass panel and depressurising the sample. d) The sample is ready.}
\label{fig:sample_prep}
\end{figure}

 Our sample is composed of soda lime glass beads of 100 $\mu m$ in diameter, placed in a brick-shaped latex membrane of 8.5 cm in height,5 cm in width, 2.5 cm in depth and 0.25 mm in thickness. Before being placed in the membrane, the beads are dried for 1h at 60°C under a pressure of 100 Pa, allowing us to control their humidity. 
 
 The preparation is depicted in fig \ref{fig:sample_prep}. First, the beads are poured into the membrane which is placed in a plastic mold to ensure its shape remains constant. We have ensured that the mass of beads poured into the membrane remains constant around 180g. This gives us a packing fraction close to the random close packing fraction (estimated at $\approx 0.63$). Then to help set the confining pressure, the top surface of the membrane is coated with a thin layer ($\approx 2 $ mm) of vacuum grease before placing the glass panel (see fig.~\ref{fig:sample_prep}.c). By activating a pump, we set the difference of pressure with the outside of the sample at $\Delta P=30 kPa$. The sample consequently becomes rigid and can now be placed in the Instron (see fig.~\ref{fig:sample_prep}.d) where the confinement pressure will remain constant and the volume will be deformed at constant velocity. Thanks to the mounting system, the sample's position is adjusted so that it sits on a fixed bottom plate. Once the back plate is placed in contact to the sample, the loading plate is slowly lowered so that it barely touches the sample. This can be monitored with the Instron's force sensor.

%%%%%%%%%%%%%%%%%%%%%%%%%%%%%%%%%%%%%%%%%%%%%%%%%%%%%%%%%%%%%%%%%%%%%%%%%%%%%%%%%%%%%%%%%%%%%%%
\subsubsection{Macroscopic behaviour of the sample}

\begin{figure*}[ht!]
\centering
\includegraphics[width=0.7\paperwidth]{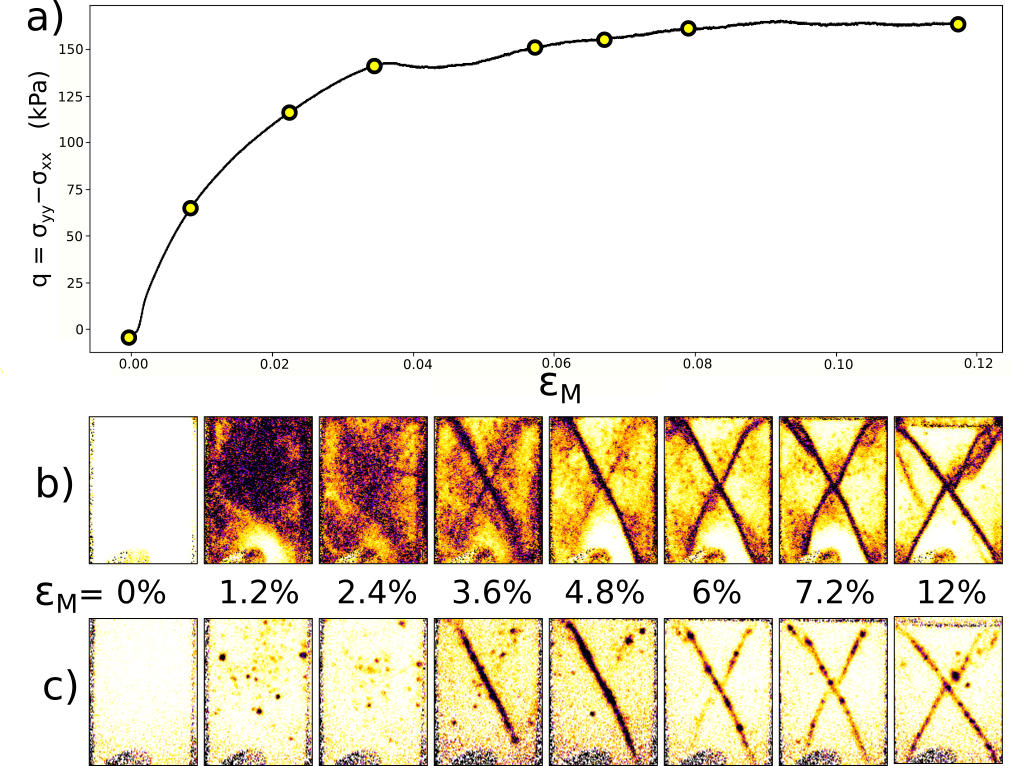}
 \caption{ a) Loading curve during an experiment. Symbols are the points corresponding to correlation maps displayed below. b) Correlation maps obtained for $\delta \varepsilon_M=5.9\times 10^{-5}$. c) Correlation maps obtained for $\delta \varepsilon_M=5.9\times 10^{-7}$.}\label{fig:global_behaviour}
\end{figure*}

We start the experiment with equal amounts of stress on the horizontal and vertical faces of our sample $\sigma_{xx}=\sigma_{yy}=\Delta P$. The sample is compressed at constant velocity $dy_{p}/dt=-1.~\mu m.s^{-1}$, where $y_{p}$ is the position of the top moving part. We note $\varepsilon_M=-(dy_{p}/dt)\times(t-t_0)/L_0$ the macroscopic deformation of the sample, where $L_0=85 mm$ is the initial height of the sample, $t_0$ is the time at which the compression force begins to increases. The evolution of $q=\sigma_{yy}-\sigma_{xx}$ as a function of the macroscopic deformation $\varepsilon_M$ is plotted on fig.~\ref{fig:global_behaviour}.a. An increase of the deviatoric stress is observed during the first 3-4\% of deformation and is followed by deformation at constant $q$. This behaviour is in agreement with numerous experiments on granular material already described in the literature. Those studies show that the first phase is associated with a plastic deformation which spreads out on the whole sample and the second phase to a localised deformation which occurs along shear bands. This will be confirmed by the measurement of the local deformation.

%%%%%%%%%%%%%%%%%%%%%%%%%%%%%%%%%%%%%%%%%%%%%%%%%%%%%%%%%%%%%%%%%%%%%%%%%%%%%%%%%%%%%%%%%%%%%%%
\subsection{Local deformation}

\subsubsection{Maps of local deformation}

The maps of local deformations are plotted on fig.\ref{fig:global_behaviour}.b-c. Those maps are calculated for some values of the macroscopic deformation $\varepsilon_M$ which are marked by circles on fig.\ref{fig:global_behaviour}.a. Each set of maps are calculated for an increment of deformation $\delta \varepsilon_M$. This means that the two states $1$ and $2$ involved in the calculus of the correlation function $g_I=(M,1,2)$ defined in eq.~\eqref{eq:defgI}, corresponds to global deformations $\varepsilon_M$ and $\varepsilon_M+\delta \varepsilon_M$. 

The set of images on fig.\ref{fig:global_behaviour}.b is calculated with $\delta \varepsilon_M=5.9\times 10^{-5}$. This corresponds to the local deformation "integrated" on a macroscopic deformation of $\delta \varepsilon_M$. We see that the local deformation is spread relatively homogeneously inside the sample before the stress plateau ($\varepsilon_M < 3\%$). On the stress plateau ($\varepsilon_M > 4\%$) the local deformation is mainly concentrated within two shear bands.

The set of images on fig.\ref{fig:global_behaviour}.c is calculated with $\delta \varepsilon_M=5.9\times 10^{-7}$. This corresponds to the instantaneous local deformation (more precisely integrated on a very small macroscopic deformation $\delta \varepsilon_M$). The local deformation which appears relatively homogeneous before the stress plateau is indeed very heterogeneous.

%%%%%%%%%%%%%%%%%%%%%%%%%%%%%%%%%%%%%%%%%%%%%%%%%%%%%%%%%%%
\subsubsection{Localised plastic event.}

\begin{figure}
\centering
\includegraphics[width=\columnwidth]{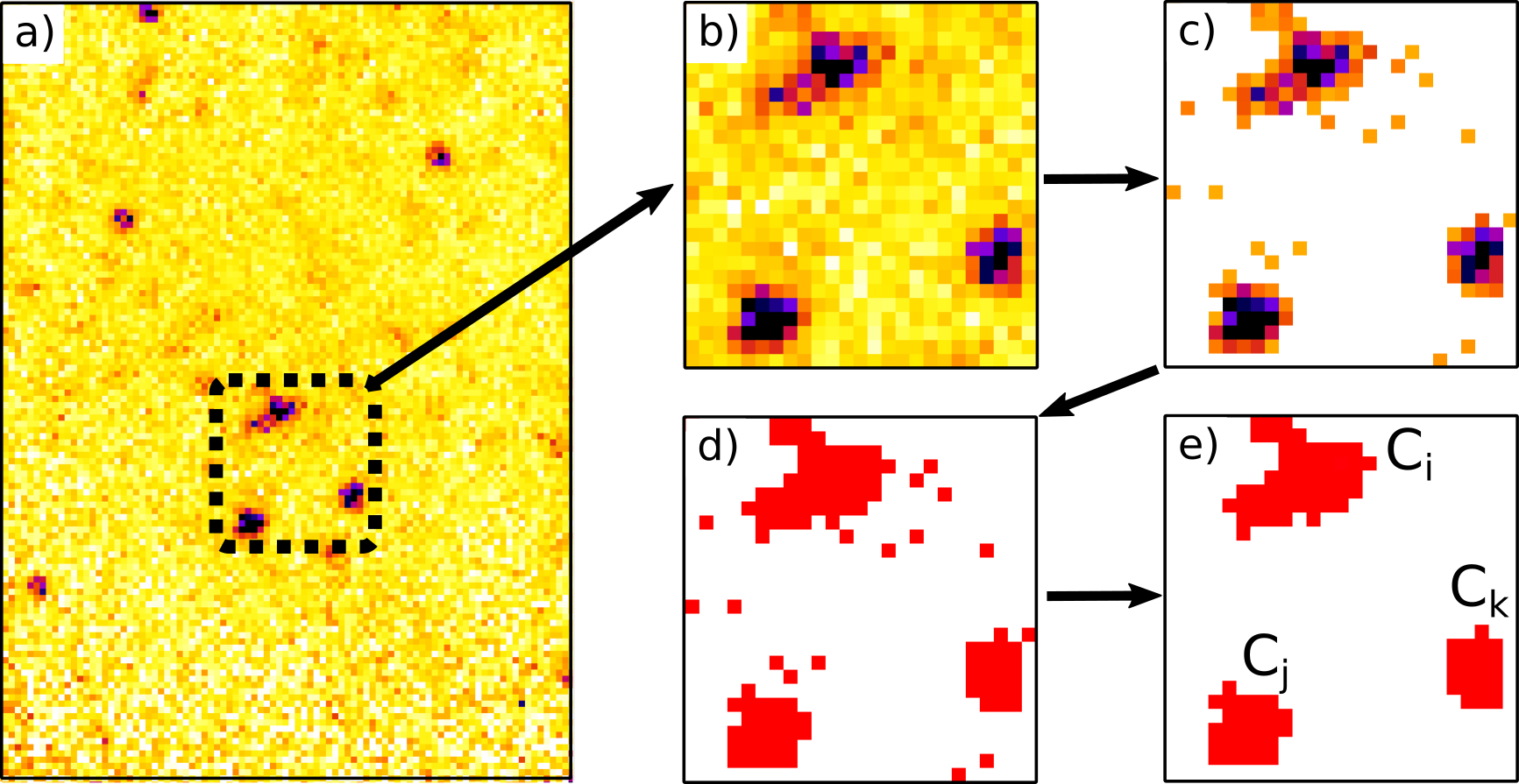}
\caption{Threshold process isolating connected sets $C_{i,j,k}$ from the background using a Connected Component Labelling algorithm.}
\label{fig:threshold_process}
\end{figure}

We define discrete "events" from an image such as fig.\ref{fig:threshold_process}.a in the following way: First, a threshold mask is applied to the image, ie all pixels with $\varepsilon \le \varepsilon_{thr}$ are set to $\varepsilon=0$ (fig.\ref{fig:threshold_process}.c). The choice of the value for $\varepsilon_{thr}$ is detailed below. In a second step, the connected zones of pixels with $\varepsilon \neq 0$ are determined using a typical Connected Component Labelling algorithm (CCL) ~\cite{CCL.article}(fig.\ref{fig:threshold_process}.d). We remove all zones that do not extend on at least $2$ Metapixels (fig.\ref{fig:threshold_process}.e). As output, we obtained a catalogue of events labelled with integer $i$, with their surfaces in metapixel $S_i=\sum_{M\in C_i}$ (${M\in C_i}$ means metapixel $M$ is in the connected set $C_i$), a mean position $(x_i,y_i)$, with $x_i \eqdef (1/S_i)\sum_{M\in C_i}x(M)$ (idem for $yi$), and a mean deformation $\varepsilon_i \eqdef (1/S_i)\sum_{M\in C_i}\varepsilon(M)$.

\begin{figure}[hpt!]
\centering
\includegraphics[width=\columnwidth]{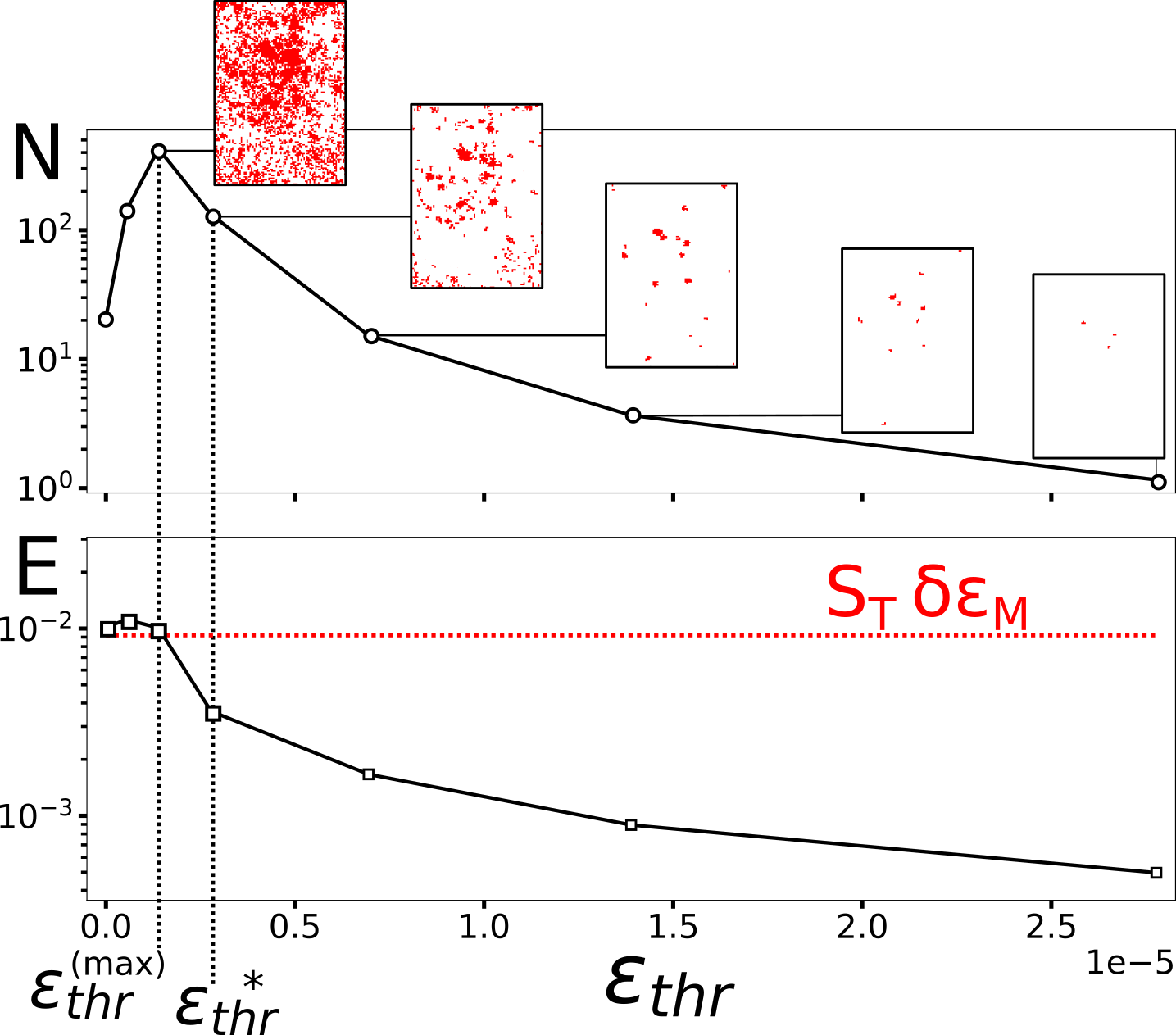}
\caption{a) Mean number of events per image N as a function of the threshold value $\varepsilon_{thr}$. b) (Symbol): mean surface-integrated deformation $E$ contained in the discrete events extracted from a single image as a function of the threshold $\varepsilon_{thr}$.  (Red dashed line): surface-integrated macroscopic deformation applied to the sample between two images. Black lines connecting symbols are  guidelines.}
\label{fig:threshold}
\end{figure}

The number of detected events as a function of the threshold $\varepsilon_{thr}$ is shown in fig.\ref{fig:threshold}.a. For small $\varepsilon_{thr}$ the CCL algorithm only detects a single event spanning the entire sample. The separation of events from their neighbours requires a higher threshold. On the other side, if $\varepsilon_{thr}$ is too high, no events will be detected. We take a value of $\varepsilon_{thr}^*$ which is twice the value of the threshold for which the number of detected events is maximal $\varepsilon_{thr}^{(max)}$ (see fig.\ref{fig:threshold}).

It should be noted that the precise value of $\varepsilon_{thr}^*$ has some effects on the measured deformations. For a given threshold $\varepsilon_{thr}$, we define:
\begin{equation}
E(\varepsilon_{thr})=\sum_i \varepsilon_i S_i-\sum_i \varepsilon_i^{n} S_i^{n}
\end{equation}
where $\sum_i \varepsilon_i S_i$ is the surface-integrated deformation contained in all the discrete events on one image, and $\sum_{i'} \varepsilon_{i'}^{n} S_{i'}^{n}$ is the surface-integrated deformation contained in all the discrete events on one image of a static object. $E(\varepsilon_{thr})$ represents the surface-integrated deformation contained in all the discrete events on one image. The variation of $E(\varepsilon_{thr})$ are plotted on fig.~\ref{fig:threshold}.b. We see that  $E(\varepsilon_{thr}^*)\sim 0.36 \times E(\varepsilon_{thr}\to 0)$. This means that for our threshold, typically $36\%$ of the deformation arises as events, and the remaining part of the deformation as a background of deformation. Although there is some arbitrary in the choice of $\varepsilon_{thr}$, we will see below that the precise value of  $\varepsilon_{thr}$ has relatively small effects on the distribution sizes of those events. 

We may also compared $E(\varepsilon_{thr}\to 0)$ to the imposed deformation. Indeed, with the knowledge of $\delta \varepsilon_M$ and the surface (expressed in number of metapixels) $S_{T}$ of the sample, we may calculate an expected integrated deformation $S_{T} \times \delta \varepsilon_M =9.2 \times 10^{-3}$. This value is reported on fig.\ref{fig:threshold}.b. We observe that for threshold value smaller or equal than $\varepsilon_{thr}^{(max)}$, the measured $E \simeq \delta \varepsilon_M\times S_{T}$. This shows that our method measure the total deformation quite accurately.

It should be noted that the value of the optimum is different before and after the localisation of the deformation. Indeed, because of strain localisation in shear bands, the average value of the deformation inside shear bands is larger than the average deformation in the sample when localisation did not occur ($\varepsilon_M < 3\%$). In the following, we only present results before localisation.

%%%%%%%%%%%%%%%%%%%%%%%%%%%%%%%%%%%%%%%%%%%%%%%%%%%%%%%%%%%
\section{Results and discussion}\label{sec:5}

\begin{figure}[ht!]
\centering
\includegraphics[width=\columnwidth]{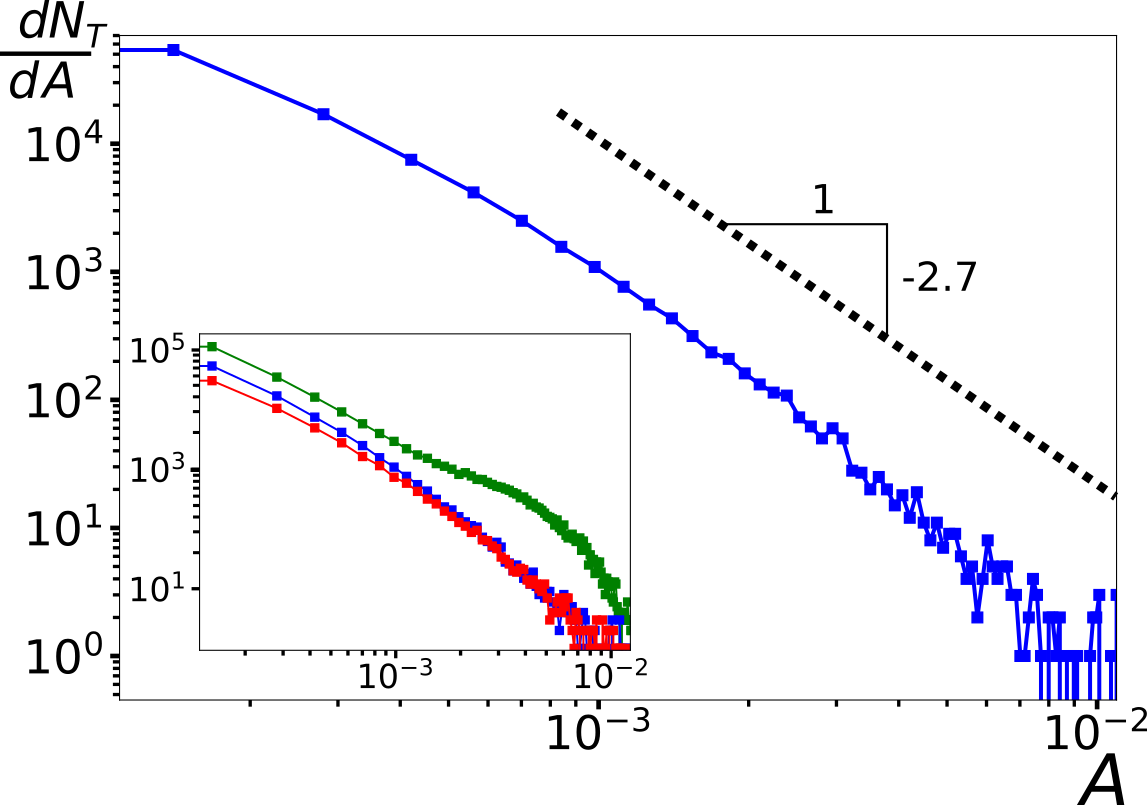}
\caption{Distribution of the events amplitudes $A_i$ (deformation rate: $\dot{\varepsilon}_M=2.4 \times 10^{-5} s^{-1}$, threshold: $\varepsilon_{thr}= \varepsilon_{thr}^{*}=2.9\times10^{-6}$).
Insert: effect of the threshold $\varepsilon_{thr}$ on the amplitude distribution. Green: $\varepsilon_{thr}=\varepsilon_{thr}^{(max)}$;  blue: $\varepsilon_{thr}=2~\varepsilon_{thr}^{(max)}=\varepsilon_{thr}^{*}$; red: $\varepsilon_{thr}=5~\varepsilon_{thr}^{(max)}$.}
\label{fig:distribution_threshold}
\end{figure}

We illustrate the possibility of our device by presenting some preliminary results on the statistics of events. In particular, the dependence of those statistics with the compression rate $\Dot{\varepsilon}_M$ is investigated. 

%%%%%%%%%%%%%%%%%%%%%%%%%%%%%%%%%%%%%%%%%%%%%%%%%%%%%%%%%%%%%%%%%%%%%%%%%%%%
\subsection{Size distribution of events}

We consider first an experiment made at $\dot{\varepsilon}_M=2.35 \times 10^{-5} s^{-1}$. For this experiment, we acquire speckles images at 40 frames per second ($\delta \varepsilon_M=5.8 \times 10^{-7} $), and we measure all the events occurring in the range  $1.8\% \le \varepsilon_M \le 2.9\%$ (ie before strain localisation). Fig.\ref{fig:distribution_threshold} shows the distribution of the amplitude $A_i \eqdef \varepsilon_i S_i$ of the events obtained for a threshold $\varepsilon_{thr}=2~\varepsilon_{thr}^{(max)}=2.8 \times 10^{-6} $. We see that those events are distributed in amplitude, and follow a power-law:
\begin{equation}
\frac{dN_T}{dA} \sim A^{-\beta}
\end{equation}
with $N_T$ the total number of events of a certain amplitude range and $\beta\approx2.7$.

It should be noticed that the size distribution is not strongly dependent on the value of $\varepsilon_{thr}$ (see inset of fig.~\ref{fig:distribution_threshold}). The effect of increasing $\varepsilon_{thr}$ is mainly to drift the size distribution towards small events, and not act as a cut-off removing events of small amplitudes and leaving the other ones unchanged. The effect of increasing $\varepsilon_{thr}$ is to remove the pixels below a given value, reducing the size of events because their rims are now below the threshold.

%%%%%%%%%%%%%%%%%%%%%%%%%%%%%%%%%%%%%%%%%%%%%%%%%%%%%%%%%%%%%%%%%%
\subsection{Deformation rate independence}

\begin{figure}[ht!]
\centering
\includegraphics[width=\columnwidth]{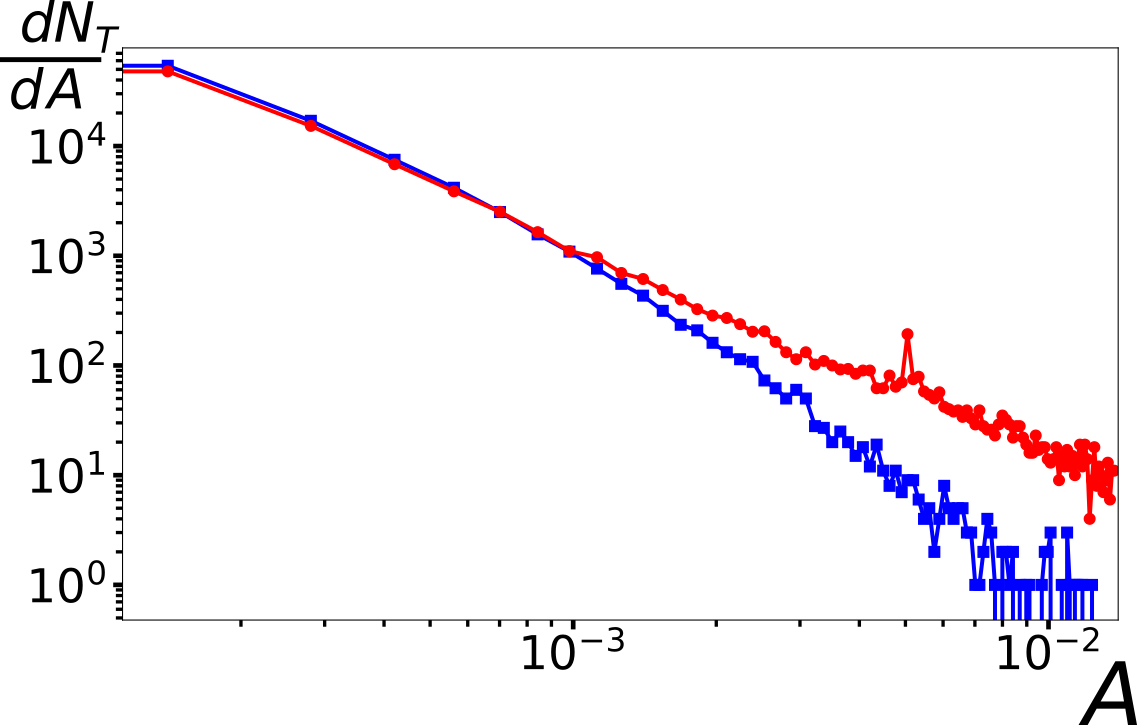}
\caption{Size distribution for $2$ different compression rates $\Dot{\varepsilon}_M$. (Red circles): $\Dot{\varepsilon}_M=2.9 \times 10^{-6} s^{-1}$, (Blue squares): $\Dot{\varepsilon}_M=2.4 \times 10^{-5} s^{-1}$. With $\delta \varepsilon_M=5.8 \times 10^{-7}$ and $\varepsilon_{thr}=2.8 \times 10 ^{-6}$.}
\label{fig:distribution_2_speeds}
\end{figure}

Due to the extended range of frame rate for speckle image acquisition, we may test the variations of the statistical properties of those events with the compression rate $\Dot{\varepsilon}_M$. For this, we perform experiments on the same material, but at 2 deformation speeds. We maintain the deformation increment $\delta \varepsilon_M$ fixed by taking a frame-rate value of $\Dot{\varepsilon}/\delta \varepsilon_M$ for image acquisition.

The figure~\ref{fig:distribution_2_speeds} shows the amplitude distributions obtained when varying the compression rate of one order of magnitude. We observe that the deformation rate does not affect significantly, at least at small size, the size distribution of events. A small increase in large events may be viewed at small velocities. It is presumably an artefact due to spurious decorrelation discussed in section~\ref{sec:low_acqui_rate}. Indeed, for $\Dot{\varepsilon}_M=2.9 \times 10^{-6} s^{-1}$, the frame rate is $2.5 s^{-1}$, and some decorrelation occurs in the determination of $g_I$ from~\eqref{eq:gI.exp}. This has the effect that two events that should be separated are indeed merged by the CCL algorithm due to this spurious decorrelation. A possible improvement may be to split those events into two separate ones by looking at the shape of such events which are non-convex. 

%%%%%%%%%%%%%%%%%%%%%%%%%%%%%%%%%%%%%%%%%%%%%%%%%%%%%%%%%%%%%%%%%%
\section{Conclusion}

In this article, we described a device specifically designed for the study of discrete plasticity events in granular media. Observation of these events requires very sensitive strain measurement (typical strain $\sim 10^{-5}$) with a spatial resolution of a few grains. We have shown that such a measurement is possible, using dynamic light scattering. The geometry used (biaxial stress test) allows a wide range of macroscopic deformation during which the sample does not localise deformation along shear bands. Particular care must be taken when measuring strain, in particular by correcting for electronic noise that may affect the measurement. Events can be defined using a standard threshold algorithm, but the choice of the minimum threshold is conditional on the presence of finite electronic noise.

The preliminary results presented in section~\ref{sec:5} show that such discrete plasticity events can be observed experimentally without ambiguity. These observations are in agreement with some recent numerical studies~\cite{Liu.2022,liu.2022a,zheng2023micromechanical}. The assertion~\cite{Bouzid.2015} that has been made in the literature, based on other numerical studies, that such discrete events do not occur in the limit of rigid systems does not apply to our experiments. This model thus appears over simplified to reproduce the experimental reality.

The quasi-independence of the statistical properties of these events with the compression speed shows that their appearance is not linked to inertial effects, nor to ageing effects of microscopic contacts. On the contrary, the onset of these events seems to be linked to the deformation imposed on the system. This is consistent with the events being triggered by local constraints in the system, which are dependent on the history of the macroscopic shock and the history of previous events. It should be noted that this independence of the statistical distribution of events has been observed in plastic events occurring after material failure, in the zone of localised deformation, i.e. in the cracking band~\cite{Houdoux2021}.

Finally, it should be noted that the existence of these globally isotropic discrete events is not contradictory with a strain field showing anisotropic correlations. In fact, it is quite legitimate to think that these discrete plastic events redistribute stress in the material. If the material behaves elastically, this redistribution of stress is given by Eshelby tensors~\cite{eshelby1957} that are anisotropic, with certain directions accumulating stress (or more precisely stress difference), and certain directions relaxing stress~\cite{Karimi2018,alexandre.2018,mcnamara.2016}. These stress variations are not visible with the strain measurement method used, because the elastic strain is too small. However, this variation in stress will bring certain zones of the material closer to their failure thresholds, and therefore increase the probability of events occurring when the load continues. Plastic deformation will therefore be correlated anisotropically {\it in the future} of the load, even if it is not correlated {\it instantaneously}. This instantaneous isotropy does not therefore contradict the delayed anistotropy observed in previous experiments~\cite{lebouil2014,lebouil.2014a,houdoux2018}. On the other hand, it indicates that the amplitude of the stress redistribution of an event must be small compared with the typical stress variation required to generate a new event. The way in which the strain correlation is constructed should provide information about these comparative amplitudes. This will be the subject of a future study.

\section*{Ethics declarations
Conflict of interest}
The authors declare that they have no known competing financial interests or personal relationships that could have appeared to influence the work reported in this paper.

\bibliography{Rennes,EarthandEnv,GM_2023,Granmat_2013,DWS,CCL}

\begin{thebibliography}{42}%
\makeatletter
\providecommand \@ifxundefined [1]{%
 \@ifx{#1\undefined}
}%
\providecommand \@ifnum [1]{%
 \ifnum #1\expandafter \@firstoftwo
 \else \expandafter \@secondoftwo
 \fi
}%
\providecommand \@ifx [1]{%
 \ifx #1\expandafter \@firstoftwo
 \else \expandafter \@secondoftwo
 \fi
}%
\providecommand \natexlab [1]{#1}%
\providecommand \enquote  [1]{``#1''}%
\providecommand \bibnamefont  [1]{#1}%
\providecommand \bibfnamefont [1]{#1}%
\providecommand \citenamefont [1]{#1}%
\providecommand \href@noop [0]{\@secondoftwo}%
\providecommand \href [0]{\begingroup \@sanitize@url \@href}%
\providecommand \@href[1]{\@@startlink{#1}\@@href}%
\providecommand \@@href[1]{\endgroup#1\@@endlink}%
\providecommand \@sanitize@url [0]{\catcode `\\12\catcode `\$12\catcode
  `\&12\catcode `\#12\catcode `\^12\catcode `\_12\catcode `\%12\relax}%
\providecommand \@@startlink[1]{}%
\providecommand \@@endlink[0]{}%
\providecommand \url  [0]{\begingroup\@sanitize@url \@url }%
\providecommand \@url [1]{\endgroup\@href {#1}{\urlprefix }}%
\providecommand \urlprefix  [0]{URL }%
\providecommand \Eprint [0]{\href }%
\providecommand \doibase [0]{https://doi.org/}%
\providecommand \selectlanguage [0]{\@gobble}%
\providecommand \bibinfo  [0]{\@secondoftwo}%
\providecommand \bibfield  [0]{\@secondoftwo}%
\providecommand \translation [1]{[#1]}%
\providecommand \BibitemOpen [0]{}%
\providecommand \bibitemStop [0]{}%
\providecommand \bibitemNoStop [0]{.\EOS\space}%
\providecommand \EOS [0]{\spacefactor3000\relax}%
\providecommand \BibitemShut  [1]{\csname bibitem#1\endcsname}%
\let\auto@bib@innerbib\@empty
%</preamble>
\bibitem [{\citenamefont {R.M.Nedderman}(1992)}]{nedderman}%
  \BibitemOpen
  \bibfield  {author} {\bibinfo {author} {\bibnamefont {R.M.Nedderman}},\
  }\href@noop {} {\emph {\bibinfo {title} {Statics and Kinematics of Granular
  Materials}}}\ (\bibinfo  {publisher} {Cambridge University Press},\ \bibinfo
  {year} {1992})\BibitemShut {NoStop}%
\bibitem [{\citenamefont {R.O.Davis}\ and\ \citenamefont
  {A.P.S.Selvadurai}(2002)}]{davis}%
  \BibitemOpen
  \bibfield  {author} {\bibinfo {author} {\bibnamefont {R.O.Davis}}\ and\
  \bibinfo {author} {\bibnamefont {A.P.S.Selvadurai}},\ }\href@noop {} {\emph
  {\bibinfo {title} {Plasticity and geomechanics}}}\ (\bibinfo  {publisher}
  {Cambridge University Press},\ \bibinfo {year} {2002})\BibitemShut {NoStop}%
\bibitem [{\citenamefont {Bruno~Andreotti}(2013)}]{andreotti.book}%
  \BibitemOpen
  \bibfield  {author} {\bibinfo {author} {\bibfnamefont {O.~P.}\ \bibnamefont
  {Bruno~Andreotti}, \bibfnamefont {Yoël~Forterre}},\ }\href@noop {} {\emph
  {\bibinfo {title} {Granular Media: Between Fluid and Solid}}}\ (\bibinfo
  {publisher} {Cambridge University Press},\ \bibinfo {year}
  {2013})\BibitemShut {NoStop}%
\bibitem [{\citenamefont {Nicolas}\ \emph {et~al.}(2018)\citenamefont
  {Nicolas}, \citenamefont {Ferrero}, \citenamefont {Martens},\ and\
  \citenamefont {Barrat}}]{alexandre.2018}%
  \BibitemOpen
  \bibfield  {author} {\bibinfo {author} {\bibfnamefont {A.}~\bibnamefont
  {Nicolas}}, \bibinfo {author} {\bibfnamefont {E.~E.}\ \bibnamefont
  {Ferrero}}, \bibinfo {author} {\bibfnamefont {K.}~\bibnamefont {Martens}},\
  and\ \bibinfo {author} {\bibfnamefont {J.-L.}\ \bibnamefont {Barrat}},\
  }\href {https://doi.org/10.1103/RevModPhys.90.045006} {\bibfield  {journal}
  {\bibinfo  {journal} {Rev. Mod. Phys.}\ }\textbf {\bibinfo {volume} {90}},\
  \bibinfo {pages} {045006} (\bibinfo {year} {2018})}\BibitemShut {NoStop}%
\bibitem [{\citenamefont {J.-P.Bardet}(1990)}]{bardet1990}%
  \BibitemOpen
  \bibfield  {author} {\bibinfo {author} {\bibnamefont {J.-P.Bardet}},\
  }\href@noop {} {\bibfield  {journal} {\bibinfo  {journal} {Computers and
  geotechnics}\ }\textbf {\bibinfo {volume} {10}},\ \bibinfo {pages} {163}
  (\bibinfo {year} {1990})}\BibitemShut {NoStop}%
\bibitem [{\citenamefont {J.Desrues}\ and\ \citenamefont
  {G.Viggiani}(2004)}]{desrues2004}%
  \BibitemOpen
  \bibfield  {author} {\bibinfo {author} {\bibnamefont {J.Desrues}}\ and\
  \bibinfo {author} {\bibnamefont {G.Viggiani}},\ }\href
  {https://doi.org/10.1002/nag.338} {\bibfield  {journal} {\bibinfo  {journal}
  {International Journal for Numerical and Analytical Methods in Geomechanics}\
  }\textbf {\bibinfo {volume} {28}},\ \bibinfo {pages} {279} (\bibinfo {year}
  {2004})}\BibitemShut {NoStop}%
\bibitem [{\citenamefont {Viggiani}\ and\ \citenamefont
  {Hall}(2012)}]{viggiani2012}%
  \BibitemOpen
  \bibfield  {author} {\bibinfo {author} {\bibfnamefont {G.}~\bibnamefont
  {Viggiani}}\ and\ \bibinfo {author} {\bibfnamefont {S.~A.}\ \bibnamefont
  {Hall}},\ }in\ \href@noop {} {\emph {\bibinfo {booktitle} {ALERT\- Doctoral
  School 2012:\- Advanced experimental techniques in geomechanics}}},\ \bibinfo
  {editor} {edited by\ \bibinfo {editor} {\bibfnamefont {G.}~\bibnamefont
  {Viggiani}}, \bibinfo {editor} {\bibfnamefont {S.}~\bibnamefont {Hall}},\
  and\ \bibinfo {editor} {\bibfnamefont {E.}~\bibnamefont {Romero}}}\ (\bibinfo
   {publisher} {ALERT Geomaterials},\ \bibinfo {year} {2012})\ pp.\ \bibinfo
  {pages} {3--67}\BibitemShut {NoStop}%
\bibitem [{\citenamefont {I.Vardoulakis}\ and\ \citenamefont
  {J.Sulem}(1995)}]{vardoulakis}%
  \BibitemOpen
  \bibfield  {author} {\bibinfo {author} {\bibnamefont {I.Vardoulakis}}\ and\
  \bibinfo {author} {\bibnamefont {J.Sulem}},\ }\href@noop {} {\emph {\bibinfo
  {title} {Bifurcation analysis in geomechanics}}}\ (\bibinfo  {publisher}
  {Blackie Academic and Professional, Glasgow, England},\ \bibinfo {year}
  {1995})\BibitemShut {NoStop}%
\bibitem [{\citenamefont {Wan}\ \emph {et~al.}(2013)\citenamefont {Wan},
  \citenamefont {Pinheiro}, \citenamefont {Daouadji}, \citenamefont {Jrad},\
  and\ \citenamefont {Darve}}]{darve.2013}%
  \BibitemOpen
  \bibfield  {author} {\bibinfo {author} {\bibfnamefont {R.}~\bibnamefont
  {Wan}}, \bibinfo {author} {\bibfnamefont {M.}~\bibnamefont {Pinheiro}},
  \bibinfo {author} {\bibfnamefont {A.}~\bibnamefont {Daouadji}}, \bibinfo
  {author} {\bibfnamefont {M.}~\bibnamefont {Jrad}},\ and\ \bibinfo {author}
  {\bibfnamefont {F.}~\bibnamefont {Darve}},\ }\href
  {https://doi.org/https://doi.org/10.1002/nag.2085} {\bibfield  {journal}
  {\bibinfo  {journal} {International Journal for Numerical and Analytical
  Methods in Geomechanics}\ }\textbf {\bibinfo {volume} {37}},\ \bibinfo
  {pages} {1292} (\bibinfo {year} {2013})},\ \Eprint
  {https://arxiv.org/abs/https://onlinelibrary.wiley.com/doi/pdf/10.1002/nag.2085}
  {https://onlinelibrary.wiley.com/doi/pdf/10.1002/nag.2085} \BibitemShut
  {NoStop}%
\bibitem [{\citenamefont {P.B\'esuelle}\ and\ \citenamefont
  {J.W.Rudnicki}(2004)}]{besuelle2004}%
  \BibitemOpen
  \bibfield  {author} {\bibinfo {author} {\bibnamefont {P.B\'esuelle}}\ and\
  \bibinfo {author} {\bibnamefont {J.W.Rudnicki}},\ }in\ \href@noop {} {\emph
  {\bibinfo {booktitle} {Mechanics of fluid-saturated rocks}}},\ \bibinfo
  {editor} {edited by\ \bibinfo {editor} {\bibnamefont {Y.Gu\'eguen}}\ and\
  \bibinfo {editor} {\bibnamefont {M.Bout\`eca}}}\ (\bibinfo  {publisher}
  {International geophysics series 89, Elsevier},\ \bibinfo {year} {2004})\
  pp.\ \bibinfo {pages} {219--321}\BibitemShut {NoStop}%
\bibitem [{\citenamefont {Lherminier}\ \emph {et~al.}(2019)\citenamefont
  {Lherminier}, \citenamefont {Planet}, \citenamefont {dit Vehel},
  \citenamefont {Simon}, \citenamefont {Vanel}, \citenamefont {M{\aa}l{\o}y},\
  and\ \citenamefont {Ramos}}]{lherminier2019}%
  \BibitemOpen
  \bibfield  {author} {\bibinfo {author} {\bibfnamefont {S.}~\bibnamefont
  {Lherminier}}, \bibinfo {author} {\bibfnamefont {R.}~\bibnamefont {Planet}},
  \bibinfo {author} {\bibfnamefont {V.~L.}\ \bibnamefont {dit Vehel}}, \bibinfo
  {author} {\bibfnamefont {G.}~\bibnamefont {Simon}}, \bibinfo {author}
  {\bibfnamefont {L.}~\bibnamefont {Vanel}}, \bibinfo {author} {\bibfnamefont
  {K.}~\bibnamefont {M{\aa}l{\o}y}},\ and\ \bibinfo {author} {\bibfnamefont
  {O.}~\bibnamefont {Ramos}},\ }\href@noop {} {\bibfield  {journal} {\bibinfo
  {journal} {Physical Review Letters}\ }\textbf {\bibinfo {volume} {122}},\
  \bibinfo {pages} {218501} (\bibinfo {year} {2019})}\BibitemShut {NoStop}%
\bibitem [{\citenamefont {Argon}(1979)}]{argon1979}%
  \BibitemOpen
  \bibfield  {author} {\bibinfo {author} {\bibfnamefont {A.}~\bibnamefont
  {Argon}},\ }\href@noop {} {\bibfield  {journal} {\bibinfo  {journal} {Acta
  metallurgica}\ }\textbf {\bibinfo {volume} {27}},\ \bibinfo {pages} {47}
  (\bibinfo {year} {1979})}\BibitemShut {NoStop}%
\bibitem [{\citenamefont {Falk}\ and\ \citenamefont
  {Langer}(1998)}]{falk.1998}%
  \BibitemOpen
  \bibfield  {author} {\bibinfo {author} {\bibfnamefont {M.~L.}\ \bibnamefont
  {Falk}}\ and\ \bibinfo {author} {\bibfnamefont {J.~S.}\ \bibnamefont
  {Langer}},\ }\href {https://doi.org/10.1103/PhysRevE.57.7192} {\bibfield
  {journal} {\bibinfo  {journal} {Phys. Rev. E}\ }\textbf {\bibinfo {volume}
  {57}},\ \bibinfo {pages} {7192} (\bibinfo {year} {1998})}\BibitemShut
  {NoStop}%
\bibitem [{\citenamefont {Picard}\ \emph {et~al.}(2004)\citenamefont {Picard},
  \citenamefont {Ajdari}, \citenamefont {Lequeux},\ and\ \citenamefont
  {Bocquet}}]{picard2004}%
  \BibitemOpen
  \bibfield  {author} {\bibinfo {author} {\bibfnamefont {G.}~\bibnamefont
  {Picard}}, \bibinfo {author} {\bibfnamefont {A.}~\bibnamefont {Ajdari}},
  \bibinfo {author} {\bibfnamefont {F.}~\bibnamefont {Lequeux}},\ and\ \bibinfo
  {author} {\bibfnamefont {L.}~\bibnamefont {Bocquet}},\ }\href@noop {}
  {\bibfield  {journal} {\bibinfo  {journal} {The European Physical Journal E}\
  }\textbf {\bibinfo {volume} {15}},\ \bibinfo {pages} {371} (\bibinfo {year}
  {2004})}\BibitemShut {NoStop}%
\bibitem [{\citenamefont {Bocquet}\ \emph {et~al.}(2009)\citenamefont
  {Bocquet}, \citenamefont {Colin},\ and\ \citenamefont
  {Ajdari}}]{bocquet.2009}%
  \BibitemOpen
  \bibfield  {author} {\bibinfo {author} {\bibfnamefont {L.}~\bibnamefont
  {Bocquet}}, \bibinfo {author} {\bibfnamefont {A.}~\bibnamefont {Colin}},\
  and\ \bibinfo {author} {\bibfnamefont {A.}~\bibnamefont {Ajdari}},\ }\href
  {https://doi.org/10.1103/PhysRevLett.103.036001} {\bibfield  {journal}
  {\bibinfo  {journal} {Phys. Rev. Lett.}\ }\textbf {\bibinfo {volume} {103}},\
  \bibinfo {pages} {036001} (\bibinfo {year} {2009})}\BibitemShut {NoStop}%
\bibitem [{\citenamefont {Goyon}\ \emph {et~al.}(2008)\citenamefont {Goyon},
  \citenamefont {Colin}, \citenamefont {Ovarlez}, \citenamefont {Ajdari},\ and\
  \citenamefont {Bocquet}}]{Goyon2008}%
  \BibitemOpen
  \bibfield  {author} {\bibinfo {author} {\bibfnamefont {J.}~\bibnamefont
  {Goyon}}, \bibinfo {author} {\bibfnamefont {A.}~\bibnamefont {Colin}},
  \bibinfo {author} {\bibfnamefont {G.}~\bibnamefont {Ovarlez}}, \bibinfo
  {author} {\bibfnamefont {A.}~\bibnamefont {Ajdari}},\ and\ \bibinfo {author}
  {\bibfnamefont {L.}~\bibnamefont {Bocquet}},\ }\href
  {https://doi.org/10.1038/nature07026} {\bibfield  {journal} {\bibinfo
  {journal} {Nature}\ }\textbf {\bibinfo {volume} {454}},\ \bibinfo {pages}
  {84} (\bibinfo {year} {2008})}\BibitemShut {NoStop}%
\bibitem [{\citenamefont {Lin}\ \emph {et~al.}(2015)\citenamefont {Lin},
  \citenamefont {Gueudr\'e}, \citenamefont {Rosso},\ and\ \citenamefont
  {Wyart}}]{lin.2015}%
  \BibitemOpen
  \bibfield  {author} {\bibinfo {author} {\bibfnamefont {J.}~\bibnamefont
  {Lin}}, \bibinfo {author} {\bibfnamefont {T.}~\bibnamefont {Gueudr\'e}},
  \bibinfo {author} {\bibfnamefont {A.}~\bibnamefont {Rosso}},\ and\ \bibinfo
  {author} {\bibfnamefont {M.}~\bibnamefont {Wyart}},\ }\href
  {https://doi.org/10.1103/PhysRevLett.115.168001} {\bibfield  {journal}
  {\bibinfo  {journal} {Phys. Rev. Lett.}\ }\textbf {\bibinfo {volume} {115}},\
  \bibinfo {pages} {168001} (\bibinfo {year} {2015})}\BibitemShut {NoStop}%
\bibitem [{\citenamefont {Bouzid}\ \emph {et~al.}(2015)\citenamefont {Bouzid},
  \citenamefont {Izzet}, \citenamefont {Trulsson}, \citenamefont {Cl{\'e}ment},
  \citenamefont {Claudin},\ and\ \citenamefont {Andreotti}}]{Bouzid.2015}%
  \BibitemOpen
  \bibfield  {author} {\bibinfo {author} {\bibfnamefont {M.}~\bibnamefont
  {Bouzid}}, \bibinfo {author} {\bibfnamefont {A.}~\bibnamefont {Izzet}},
  \bibinfo {author} {\bibfnamefont {M.}~\bibnamefont {Trulsson}}, \bibinfo
  {author} {\bibfnamefont {E.}~\bibnamefont {Cl{\'e}ment}}, \bibinfo {author}
  {\bibfnamefont {P.}~\bibnamefont {Claudin}},\ and\ \bibinfo {author}
  {\bibfnamefont {B.}~\bibnamefont {Andreotti}},\ }\href
  {https://doi.org/10.1140/epje/i2015-15125-1} {\bibfield  {journal} {\bibinfo
  {journal} {The European Physical Journal E}\ }\textbf {\bibinfo {volume}
  {38}},\ \bibinfo {pages} {125} (\bibinfo {year} {2015})}\BibitemShut
  {NoStop}%
\bibitem [{\citenamefont {Liu}\ \emph {et~al.}(2022{\natexlab{a}})\citenamefont
  {Liu}, \citenamefont {Wautier}, \citenamefont {Zhou}, \citenamefont {Nicot},\
  and\ \citenamefont {Darve}}]{Liu.2022}%
  \BibitemOpen
  \bibfield  {author} {\bibinfo {author} {\bibfnamefont {J.}~\bibnamefont
  {Liu}}, \bibinfo {author} {\bibfnamefont {A.}~\bibnamefont {Wautier}},
  \bibinfo {author} {\bibfnamefont {W.}~\bibnamefont {Zhou}}, \bibinfo {author}
  {\bibfnamefont {F.}~\bibnamefont {Nicot}},\ and\ \bibinfo {author}
  {\bibfnamefont {F.}~\bibnamefont {Darve}},\ }\href
  {https://doi.org/10.1007/s10035-022-01258-y} {\bibfield  {journal} {\bibinfo
  {journal} {Granular Matter}\ }\textbf {\bibinfo {volume} {24}},\ \bibinfo
  {pages} {119} (\bibinfo {year} {2022}{\natexlab{a}})}\BibitemShut {NoStop}%
\bibitem [{\citenamefont {Liu}\ \emph {et~al.}(2022{\natexlab{b}})\citenamefont
  {Liu}, \citenamefont {Wautier}, \citenamefont {Nicot}, \citenamefont
  {Darve},\ and\ \citenamefont {Zhou}}]{liu.2022a}%
  \BibitemOpen
  \bibfield  {author} {\bibinfo {author} {\bibfnamefont {J.}~\bibnamefont
  {Liu}}, \bibinfo {author} {\bibfnamefont {A.}~\bibnamefont {Wautier}},
  \bibinfo {author} {\bibfnamefont {F.}~\bibnamefont {Nicot}}, \bibinfo
  {author} {\bibfnamefont {F.}~\bibnamefont {Darve}},\ and\ \bibinfo {author}
  {\bibfnamefont {W.}~\bibnamefont {Zhou}},\ }\href
  {https://doi.org/https://doi.org/10.1016/j.ijsolstr.2022.111835} {\bibfield
  {journal} {\bibinfo  {journal} {International Journal of Solids and
  Structures}\ }\textbf {\bibinfo {volume} {252}},\ \bibinfo {pages} {111835}
  (\bibinfo {year} {2022}{\natexlab{b}})}\BibitemShut {NoStop}%
\bibitem [{\citenamefont {Zheng}\ \emph {et~al.}(2023)\citenamefont {Zheng},
  \citenamefont {Sun},\ and\ \citenamefont {Zhang}}]{zheng2023micromechanical}%
  \BibitemOpen
  \bibfield  {author} {\bibinfo {author} {\bibfnamefont {J.}~\bibnamefont
  {Zheng}}, \bibinfo {author} {\bibfnamefont {A.}~\bibnamefont {Sun}},\ and\
  \bibinfo {author} {\bibfnamefont {J.}~\bibnamefont {Zhang}},\ }\href@noop {}
  {\bibinfo {title} {Micromechanical measurements of local plastic events in
  granular materials}} (\bibinfo {year} {2023}),\ \Eprint
  {https://arxiv.org/abs/2305.03218} {arXiv:2305.03218 [cond-mat.soft]}
  \BibitemShut {NoStop}%
\bibitem [{\citenamefont {Ma}\ \emph {et~al.}(2021)\citenamefont {Ma},
  \citenamefont {Zou}, \citenamefont {Chen}, \citenamefont {Tang},
  \citenamefont {tat Ng},\ and\ \citenamefont {Zhou}}]{ma.2021}%
  \BibitemOpen
  \bibfield  {author} {\bibinfo {author} {\bibfnamefont {G.}~\bibnamefont
  {Ma}}, \bibinfo {author} {\bibfnamefont {Y.}~\bibnamefont {Zou}}, \bibinfo
  {author} {\bibfnamefont {Y.}~\bibnamefont {Chen}}, \bibinfo {author}
  {\bibfnamefont {L.}~\bibnamefont {Tang}}, \bibinfo {author} {\bibfnamefont
  {T.}~\bibnamefont {tat Ng}},\ and\ \bibinfo {author} {\bibfnamefont
  {W.}~\bibnamefont {Zhou}},\ }\href
  {https://doi.org/https://doi.org/10.1016/j.powtec.2020.09.053} {\bibfield
  {journal} {\bibinfo  {journal} {Powder Technology}\ }\textbf {\bibinfo
  {volume} {378}},\ \bibinfo {pages} {263} (\bibinfo {year}
  {2021})}\BibitemShut {NoStop}%
\bibitem [{\citenamefont {Le~Bouil}\ \emph
  {et~al.}(2014{\natexlab{a}})\citenamefont {Le~Bouil}, \citenamefont {Amon},
  \citenamefont {McNamara},\ and\ \citenamefont {Crassous}}]{lebouil2014}%
  \BibitemOpen
  \bibfield  {author} {\bibinfo {author} {\bibfnamefont {A.}~\bibnamefont
  {Le~Bouil}}, \bibinfo {author} {\bibfnamefont {A.}~\bibnamefont {Amon}},
  \bibinfo {author} {\bibfnamefont {S.}~\bibnamefont {McNamara}},\ and\
  \bibinfo {author} {\bibfnamefont {J.}~\bibnamefont {Crassous}},\ }\href
  {https://doi.org/10.1103/PhysRevLett.112.246001} {\bibfield  {journal}
  {\bibinfo  {journal} {Phys. Rev. Lett.}\ }\textbf {\bibinfo {volume} {112}},\
  \bibinfo {pages} {246001} (\bibinfo {year} {2014}{\natexlab{a}})}\BibitemShut
  {NoStop}%
\bibitem [{\citenamefont {Houdoux}\ \emph {et~al.}(2018)\citenamefont
  {Houdoux}, \citenamefont {Nguyen}, \citenamefont {Amon},\ and\ \citenamefont
  {Crassous}}]{houdoux2018}%
  \BibitemOpen
  \bibfield  {author} {\bibinfo {author} {\bibfnamefont {D.}~\bibnamefont
  {Houdoux}}, \bibinfo {author} {\bibfnamefont {T.~B.}\ \bibnamefont {Nguyen}},
  \bibinfo {author} {\bibfnamefont {A.}~\bibnamefont {Amon}},\ and\ \bibinfo
  {author} {\bibfnamefont {J.}~\bibnamefont {Crassous}},\ }\href
  {https://doi.org/10.1103/PhysRevE.98.022905} {\bibfield  {journal} {\bibinfo
  {journal} {Phys. Rev. E}\ }\textbf {\bibinfo {volume} {98}},\ \bibinfo
  {pages} {022905} (\bibinfo {year} {2018})}\BibitemShut {NoStop}%
\bibitem [{\citenamefont {M.R.Kuhn}(1999)}]{kuhn1999}%
  \BibitemOpen
  \bibfield  {author} {\bibinfo {author} {\bibnamefont {M.R.Kuhn}},\ }\href
  {https://doi.org/10.1016/S0167-6636(99)00010-1} {\bibfield  {journal}
  {\bibinfo  {journal} {Mechanics of Materials}\ }\textbf {\bibinfo {volume}
  {31}},\ \bibinfo {pages} {407 } (\bibinfo {year} {1999})}\BibitemShut
  {NoStop}%
\bibitem [{\citenamefont {Guo}\ and\ \citenamefont {Zhao}(2014)}]{guo2014}%
  \BibitemOpen
  \bibfield  {author} {\bibinfo {author} {\bibfnamefont {N.}~\bibnamefont
  {Guo}}\ and\ \bibinfo {author} {\bibfnamefont {J.}~\bibnamefont {Zhao}},\
  }\href {https://doi.org/10.1103/PhysRevE.89.042208} {\bibfield  {journal}
  {\bibinfo  {journal} {Phys. Rev. E}\ }\textbf {\bibinfo {volume} {89}},\
  \bibinfo {pages} {042208} (\bibinfo {year} {2014})}\BibitemShut {NoStop}%
\bibitem [{\citenamefont {McNamara}\ \emph {et~al.}(2016)\citenamefont
  {McNamara}, \citenamefont {Crassous},\ and\ \citenamefont
  {Amon}}]{mcnamara.2016}%
  \BibitemOpen
  \bibfield  {author} {\bibinfo {author} {\bibfnamefont {S.}~\bibnamefont
  {McNamara}}, \bibinfo {author} {\bibfnamefont {J.}~\bibnamefont {Crassous}},\
  and\ \bibinfo {author} {\bibfnamefont {A.}~\bibnamefont {Amon}},\ }\href
  {https://doi.org/10.1103/PhysRevE.94.022907} {\bibfield  {journal} {\bibinfo
  {journal} {Phys. Rev. E}\ }\textbf {\bibinfo {volume} {94}},\ \bibinfo
  {pages} {022907} (\bibinfo {year} {2016})}\BibitemShut {NoStop}%
\bibitem [{\citenamefont {Darve}\ \emph {et~al.}(2021)\citenamefont {Darve},
  \citenamefont {Nicot}, \citenamefont {Wautier},\ and\ \citenamefont
  {Liu}}]{darve2021}%
  \BibitemOpen
  \bibfield  {author} {\bibinfo {author} {\bibfnamefont {F.}~\bibnamefont
  {Darve}}, \bibinfo {author} {\bibfnamefont {F.}~\bibnamefont {Nicot}},
  \bibinfo {author} {\bibfnamefont {A.}~\bibnamefont {Wautier}},\ and\ \bibinfo
  {author} {\bibfnamefont {J.}~\bibnamefont {Liu}},\ }\href
  {https://doi.org/https://doi.org/10.1016/j.mechrescom.2020.103603} {\bibfield
   {journal} {\bibinfo  {journal} {Mechanics Research Communications}\ }\textbf
  {\bibinfo {volume} {114}},\ \bibinfo {pages} {103603} (\bibinfo {year}
  {2021})},\ \bibinfo {note} {special Issue in Honor of Prof. N. D.
  Cristescu}\BibitemShut {NoStop}%
\bibitem [{\citenamefont {Wang}\ \emph {et~al.}(2022)\citenamefont {Wang},
  \citenamefont {Lu}, \citenamefont {Lu}, \citenamefont {Huo}, \citenamefont
  {Wang}, \citenamefont {Wang}, \citenamefont {Dai},\ and\ \citenamefont
  {Jiang}}]{Wang2022}%
  \BibitemOpen
  \bibfield  {author} {\bibinfo {author} {\bibfnamefont {X.~J.}\ \bibnamefont
  {Wang}}, \bibinfo {author} {\bibfnamefont {Y.~Z.}\ \bibnamefont {Lu}},
  \bibinfo {author} {\bibfnamefont {X.}~\bibnamefont {Lu}}, \bibinfo {author}
  {\bibfnamefont {J.~T.}\ \bibnamefont {Huo}}, \bibinfo {author} {\bibfnamefont
  {Y.~J.}\ \bibnamefont {Wang}}, \bibinfo {author} {\bibfnamefont {W.~H.}\
  \bibnamefont {Wang}}, \bibinfo {author} {\bibfnamefont {L.~H.}\ \bibnamefont
  {Dai}},\ and\ \bibinfo {author} {\bibfnamefont {M.~Q.}\ \bibnamefont
  {Jiang}},\ }\href {https://doi.org/10.1103/PhysRevE.105.045003} {\bibfield
  {journal} {\bibinfo  {journal} {Phys. Rev. E}\ }\textbf {\bibinfo {volume}
  {105}},\ \bibinfo {pages} {045003} (\bibinfo {year} {2022})}\BibitemShut
  {NoStop}%
\bibitem [{\citenamefont {Eshelby}\ and\ \citenamefont
  {Peierls}(1957)}]{eshelby1957}%
  \BibitemOpen
  \bibfield  {author} {\bibinfo {author} {\bibfnamefont {J.~D.}\ \bibnamefont
  {Eshelby}}\ and\ \bibinfo {author} {\bibfnamefont {R.~E.}\ \bibnamefont
  {Peierls}},\ }\href {https://doi.org/10.1098/rspa.1957.0133} {\bibfield
  {journal} {\bibinfo  {journal} {Proceedings of the Royal Society of London.
  Series A. Mathematical and Physical Sciences}\ }\textbf {\bibinfo {volume}
  {241}},\ \bibinfo {pages} {376} (\bibinfo {year} {1957})},\ \Eprint
  {https://arxiv.org/abs/https://royalsocietypublishing.org/doi/pdf/10.1098/rspa.1957.0133}
  {https://royalsocietypublishing.org/doi/pdf/10.1098/rspa.1957.0133}
  \BibitemShut {NoStop}%
\bibitem [{\citenamefont {Karimi}\ and\ \citenamefont
  {Barrat}(2018)}]{Karimi2018}%
  \BibitemOpen
  \bibfield  {author} {\bibinfo {author} {\bibfnamefont {K.}~\bibnamefont
  {Karimi}}\ and\ \bibinfo {author} {\bibfnamefont {J.-L.}\ \bibnamefont
  {Barrat}},\ }\href {https://doi.org/10.1038/s41598-018-22310-z} {\bibfield
  {journal} {\bibinfo  {journal} {Scientific Reports}\ }\textbf {\bibinfo
  {volume} {8}},\ \bibinfo {pages} {4021} (\bibinfo {year} {2018})}\BibitemShut
  {NoStop}%
\bibitem [{\citenamefont {Le~Bouil}\ \emph
  {et~al.}(2014{\natexlab{b}})\citenamefont {Le~Bouil}, \citenamefont {Amon},
  \citenamefont {Sangleboeuf}, \citenamefont {Orain}, \citenamefont
  {B{\'e}suelle}, \citenamefont {Viggiani}, \citenamefont {Chasle},\ and\
  \citenamefont {Crassous}}]{lebouil.2014a}%
  \BibitemOpen
  \bibfield  {author} {\bibinfo {author} {\bibfnamefont {A.}~\bibnamefont
  {Le~Bouil}}, \bibinfo {author} {\bibfnamefont {A.}~\bibnamefont {Amon}},
  \bibinfo {author} {\bibfnamefont {J.-C.}\ \bibnamefont {Sangleboeuf}},
  \bibinfo {author} {\bibfnamefont {H.}~\bibnamefont {Orain}}, \bibinfo
  {author} {\bibfnamefont {P.}~\bibnamefont {B{\'e}suelle}}, \bibinfo {author}
  {\bibfnamefont {G.}~\bibnamefont {Viggiani}}, \bibinfo {author}
  {\bibfnamefont {P.}~\bibnamefont {Chasle}},\ and\ \bibinfo {author}
  {\bibfnamefont {J.}~\bibnamefont {Crassous}},\ }\href
  {https://doi.org/10.1007/s10035-013-0477-x} {\bibfield  {journal} {\bibinfo
  {journal} {Granular Matter}\ }\textbf {\bibinfo {volume} {16}},\ \bibinfo
  {pages} {1} (\bibinfo {year} {2014}{\natexlab{b}})}\BibitemShut {NoStop}%
\bibitem [{\citenamefont {Djaoui}\ and\ \citenamefont
  {Crassous}(2005)}]{djaoui.2005}%
  \BibitemOpen
  \bibfield  {author} {\bibinfo {author} {\bibfnamefont {L.}~\bibnamefont
  {Djaoui}}\ and\ \bibinfo {author} {\bibfnamefont {J.}~\bibnamefont
  {Crassous}},\ }\href {https://doi.org/10.1007/s10035-005-0210-5} {\bibfield
  {journal} {\bibinfo  {journal} {Granular Matter}\ }\textbf {\bibinfo {volume}
  {7}},\ \bibinfo {pages} {185} (\bibinfo {year} {2005})}\BibitemShut {NoStop}%
\bibitem [{\citenamefont {Crassous}(2007)}]{crassous.2007}%
  \BibitemOpen
  \bibfield  {author} {\bibinfo {author} {\bibfnamefont {J.}~\bibnamefont
  {Crassous}},\ }\href@noop {} {\bibfield  {journal} {\bibinfo  {journal} {Eur.
  Phys. J. E}\ }\textbf {\bibinfo {volume} {23}},\ \bibinfo {pages} {145}
  (\bibinfo {year} {2007})}\BibitemShut {NoStop}%
\bibitem [{\citenamefont {Erpelding}\ \emph {et~al.}(2010)\citenamefont
  {Erpelding}, \citenamefont {Guillermic}, \citenamefont {Dollet},
  \citenamefont {Saint-Jalmes},\ and\ \citenamefont
  {Crassous}}]{erpelding.2010}%
  \BibitemOpen
  \bibfield  {author} {\bibinfo {author} {\bibfnamefont {M.}~\bibnamefont
  {Erpelding}}, \bibinfo {author} {\bibfnamefont {R.~M.}\ \bibnamefont
  {Guillermic}}, \bibinfo {author} {\bibfnamefont {B.}~\bibnamefont {Dollet}},
  \bibinfo {author} {\bibfnamefont {A.}~\bibnamefont {Saint-Jalmes}},\ and\
  \bibinfo {author} {\bibfnamefont {J.}~\bibnamefont {Crassous}},\ }\href@noop
  {} {\bibfield  {journal} {\bibinfo  {journal} {Phys. Rev. E}\ }\textbf
  {\bibinfo {volume} {82}},\ \bibinfo {pages} {021409} (\bibinfo {year}
  {2010})}\BibitemShut {NoStop}%
\bibitem [{\citenamefont {Amon}\ \emph {et~al.}(2017)\citenamefont {Amon},
  \citenamefont {Mikhailovskaya},\ and\ \citenamefont {Crassous}}]{amon.2017}%
  \BibitemOpen
  \bibfield  {author} {\bibinfo {author} {\bibfnamefont {A.}~\bibnamefont
  {Amon}}, \bibinfo {author} {\bibfnamefont {A.}~\bibnamefont
  {Mikhailovskaya}},\ and\ \bibinfo {author} {\bibfnamefont {J.}~\bibnamefont
  {Crassous}},\ }\href {https://doi.org/10.1063/1.4983048} {\bibfield
  {journal} {\bibinfo  {journal} {Review of Scientific Instruments}\ }\textbf
  {\bibinfo {volume} {88}},\ \bibinfo {pages} {051804} (\bibinfo {year}
  {2017})}\BibitemShut {NoStop}%
\bibitem [{\citenamefont {{Pine, D.J.}}\ \emph {et~al.}(1990)\citenamefont
  {{Pine, D.J.}}, \citenamefont {{Weitz, D.A.}}, \citenamefont {{Zhu, J.X.}},\
  and\ \citenamefont {{Herbolzheimer, E.}}}]{pine.1990}%
  \BibitemOpen
  \bibfield  {author} {\bibinfo {author} {\bibnamefont {{Pine, D.J.}}},
  \bibinfo {author} {\bibnamefont {{Weitz, D.A.}}}, \bibinfo {author}
  {\bibnamefont {{Zhu, J.X.}}},\ and\ \bibinfo {author} {\bibnamefont
  {{Herbolzheimer, E.}}},\ }\href
  {https://doi.org/10.1051/jphys:0199000510180210100} {\bibfield  {journal}
  {\bibinfo  {journal} {J. Phys. France}\ }\textbf {\bibinfo {volume} {51}},\
  \bibinfo {pages} {2101} (\bibinfo {year} {1990})}\BibitemShut {NoStop}%
\bibitem [{\citenamefont {{D. J. Pine}and{D. A. Weitz}and{G. Maret}and{P. E.
  Wolf}and{E. Herbolzheomer}}\ and\ \citenamefont {{P. M.
  Chaikin}}(1990)}]{pine.1990b}%
  \BibitemOpen
  \bibfield  {author} {\bibinfo {author} {\bibnamefont {{D. J. Pine}and{D. A.
  Weitz}and{G. Maret}and{P. E. Wolf}and{E. Herbolzheomer}}}\ and\ \bibinfo
  {author} {\bibnamefont {{P. M. Chaikin}}},\ }in\ \href@noop {} {\emph
  {\bibinfo {booktitle} {Dynamical Correlations of Multiply Scattered
  Light}}},\ \bibinfo {editor} {edited by\ \bibinfo {editor} {\bibfnamefont
  {P.}~\bibnamefont {Sheng}}}\ (\bibinfo  {publisher} {World Scientific,
  Singapore},\ \bibinfo {year} {1990})\ pp.\ \bibinfo {pages}
  {312--372}\BibitemShut {NoStop}%
\bibitem [{\citenamefont {Viasnoff}\ \emph {et~al.}(2002)\citenamefont
  {Viasnoff}, \citenamefont {Lequeux},\ and\ \citenamefont
  {Pine}}]{viasnoff.2002}%
  \BibitemOpen
  \bibfield  {author} {\bibinfo {author} {\bibfnamefont {V.}~\bibnamefont
  {Viasnoff}}, \bibinfo {author} {\bibfnamefont {F.}~\bibnamefont {Lequeux}},\
  and\ \bibinfo {author} {\bibfnamefont {D.}~\bibnamefont {Pine}},\ }\href@noop
  {} {\bibfield  {journal} {\bibinfo  {journal} {Rev. Sci. Instrum.}\ }\textbf
  {\bibinfo {volume} {73}},\ \bibinfo {pages} {2336} (\bibinfo {year}
  {2002})}\BibitemShut {NoStop}%
\bibitem [{\citenamefont {Aime}\ \emph {et~al.}(2018)\citenamefont {Aime},
  \citenamefont {Ramos},\ and\ \citenamefont {Cipelletti}}]{aime.2018}%
  \BibitemOpen
  \bibfield  {author} {\bibinfo {author} {\bibfnamefont {S.}~\bibnamefont
  {Aime}}, \bibinfo {author} {\bibfnamefont {L.}~\bibnamefont {Ramos}},\ and\
  \bibinfo {author} {\bibfnamefont {L.}~\bibnamefont {Cipelletti}},\ }\href
  {https://doi.org/10.1073/pnas.1717403115} {\bibfield  {journal} {\bibinfo
  {journal} {Proceedings of the National Academy of Sciences}\ }\textbf
  {\bibinfo {volume} {115}},\ \bibinfo {pages} {3587} (\bibinfo {year}
  {2018})},\ \Eprint
  {https://arxiv.org/abs/https://www.pnas.org/content/115/14/3587.full.pdf}
  {https://www.pnas.org/content/115/14/3587.full.pdf} \BibitemShut {NoStop}%
\bibitem [{\citenamefont {Dillencourt}\ \emph {et~al.}(1992)\citenamefont
  {Dillencourt}, \citenamefont {Samet},\ and\ \citenamefont
  {Tamminen}}]{CCL.article}%
  \BibitemOpen
  \bibfield  {author} {\bibinfo {author} {\bibfnamefont {M.~B.}\ \bibnamefont
  {Dillencourt}}, \bibinfo {author} {\bibfnamefont {H.}~\bibnamefont {Samet}},\
  and\ \bibinfo {author} {\bibfnamefont {M.}~\bibnamefont {Tamminen}},\ }\href
  {https://doi.org/10.1145/128749.128750} {\bibfield  {journal} {\bibinfo
  {journal} {J. ACM}\ }\textbf {\bibinfo {volume} {39}},\ \bibinfo {pages}
  {253–280} (\bibinfo {year} {1992})}\BibitemShut {NoStop}%
\bibitem [{\citenamefont {Houdoux}\ \emph {et~al.}(2021)\citenamefont
  {Houdoux}, \citenamefont {Amon}, \citenamefont {Marsan}, \citenamefont
  {Weiss},\ and\ \citenamefont {Crassous}}]{Houdoux2021}%
  \BibitemOpen
  \bibfield  {author} {\bibinfo {author} {\bibfnamefont {D.}~\bibnamefont
  {Houdoux}}, \bibinfo {author} {\bibfnamefont {A.}~\bibnamefont {Amon}},
  \bibinfo {author} {\bibfnamefont {D.}~\bibnamefont {Marsan}}, \bibinfo
  {author} {\bibfnamefont {J.}~\bibnamefont {Weiss}},\ and\ \bibinfo {author}
  {\bibfnamefont {J.}~\bibnamefont {Crassous}},\ }\href
  {https://doi.org/10.1038/s43247-021-00147-1} {\bibfield  {journal} {\bibinfo
  {journal} {Communications Earth {\&} Environment}\ }\textbf {\bibinfo
  {volume} {2}},\ \bibinfo {pages} {90} (\bibinfo {year} {2021})}\BibitemShut
  {NoStop}%
\end{thebibliography}%

\end{document}